\documentclass[aps,pre,twocolumn,amsmath,amssymb,superscriptaddress,reprint,longbibliography]{revtex4-1}

\AtBeginDocument{
 \newwrite\bibnotes
 \def\bibnotesext{Notes.bib}
 \immediate\openout\bibnotes=\jobname\bibnotesext
 \immediate\write\bibnotes{@CONTROL{REVTEX41Control}}
 \immediate\write\bibnotes{@CONTROL{
 apsrev41Control,author="08",editor="1",pages="1",title="0",year="1"}}
 \if@filesw
 \immediate\write\@auxout{\string\citation{apsrev41Control}}%
 \fi
}

\usepackage{amsmath}
\usepackage{mathscinet}
\usepackage{amssymb}
\usepackage{fancyhdr}
\usepackage{emptypage} 
\usepackage[toc, title]{appendix}
\usepackage{tabularx}
\usepackage{comment}
\usepackage{graphicx}
\usepackage{multirow, array}
\usepackage[utf8]{inputenc}
\usepackage{indentfirst}
\usepackage{hyperref} % Hyperlinks.
\usepackage[]{xcolor}
\usepackage{ulem} %\sout
\hypersetup{
 colorlinks=true,
 linkcolor=violet,
 filecolor=brown, 
 urlcolor=blue,
 citecolor = olive
}
\usepackage{color}

\begin{document}
\title{Noisy kinetic-exchange opinion model with aging}
\author{Allan R. Vieira}
\address{
Department of Physics, Pontifical Catholic University of Rio de Janeiro, PUC-Rio,
Rua Marquês de São Vicente, 225, 22451-900 Rio de Janeiro, Brazil}
\author{Jaume Llabrés} 
\address{Instituto de Física Interdisciplinar y Sistemas Complejos IFISC (CSIC-UIB),
Campus Universitat de les Illes Balears, 07122 Palma de Mallorca, Spain}
\author{Ra\'ul Toral}
\address{Instituto de Física Interdisciplinar y Sistemas Complejos IFISC (CSIC-UIB),
Campus Universitat de les Illes Balears, 07122 Palma de Mallorca, Spain}
\author{Celia Anteneodo}
\address{
Department of Physics, Pontifical Catholic University of Rio de Janeiro, PUC-Rio,
Rua Marquês de São Vicente, 225, 22451-900 Rio de Janeiro, Brazil}
\address{National Institute of Science and Technology for Complex Systems, INCT-SC, Rio de Janeiro, Brazil}

\begin{abstract}
We study the critical behavior of a noisy kinetic opinion model subject to resilience to change depending on aging, defined as the time spent on the current opinion state. In this model, the opinion of each agent can take the three discrete values, the extreme ones $\pm 1$, and also the intermediate value 0, and it can evolve through kinetic exchange when interacting with another agent, or independently, by stochastic choice (noise). The probability of change by pairwise interactions depends on the age that the agent has remained in the same state, according to a given kernel. We particularly develop the cases where the probability decays either algebraically or exponentially with the age, and we also consider the anti-aging scenario where the probability increases with the age, meaning that it is more likely to change mind the longer the persistence in the current state.
For the opinion dynamics in a complete graph, we obtain analytical predictions for the critical curves of the order parameters, in perfect agreement with agent-based simulations. We observe that sufficiently weak aging (slow change of the kernel with the age) favors partial consensus with respect to the aging-insensitive scenario, while for sufficiently strong aging, as well as for anti-aging, the opposite trend is observed. 
\end{abstract}
\maketitle

\section{Introduction}
In recent years there has been an increasing interest in the application of methods coming from Statistical and Nonlinear Physics to study problems of relevance in the dynamics of social systems~\cite{Castellano:2009,SanMiguel2020b}. The main intention is to provide a ``microscopic'' explanation in terms of the individual's behaviors of observed ``macroscopic'' features, such as the emergence of consensus~\cite{Clifford:1973}, spread of innovations~\cite{Martins:2009}, global polarization~\cite{Axelrod:1997}, etc. Much of the research has employed simplified models designed to capture the essence of the phenomena under investigation while maintaining the simplicity of model components. Consequently, it is important to identify the relevant mechanisms that are able to explain a given phenomenon and include them in the model description. In the context of opinion-formation, studying the evolution of consensus in societies, it has been determined that the consideration of non-Markovian or memory effects is one of those relevant ingredients capable of modifying qualitatively the outcomes of the model~\cite{JEDRZEJEWSKI2018306,OESTEREICH2022128199}. Within the context of the current paper, we mention one of the early studies in that direction by Stark et al.~\cite{Stark2008} who introduced ``inertia'' as the property of agents of being less likely to change their state the longer they have persisted in the current one. By considering a time-dependent transition rate, these authors showed that in the context of the voter model~\cite{Clifford:1973,Holley:1975}, the slowing down of the individual's dynamics induced by inertia led, counter-intuitively, to a decrease in the time needed to reach the consensus state. Borrowing concepts from the physics literature, the property of inertia was also referred to as ``aging'' as a general term representing the influence that persistence times have on the state transitions in a system~\cite{FernandezGracia2011,Perez2016,boguna2014,abella2023ordering}. 

Subsequent research further substantiated this general trend, revealing that the introduction of aging in the agent's dynamics can not only shift the critical points of phase transitions but also alter their inherent nature. This highlights the profound impact that the temporal aspect of inertia, or aging, can have on the dynamics and critical characteristics of a system. For example, it is known that in the noisy voter model, also known as Kirman model~\cite{kirman}, where agents can hold one of two symmetrical opinions represented by $\pm1$, there is a discontinuous transition from a state of consensus (a large majority of the population holding the same opinion state) to a state of polarization (in which both opinion states are shared by approximately half of the population). The transition point of this phase transition scales as the inverse of the system size and, hence, it disappears in the thermodynamic limit in which the population size tends to infinity. Remarkably, the introduction of aging in this model has been reported to give rise to a continuous phase transition between consensus and polarization which is well defined in the thermodynamic limit, with a non-zero critical point~\cite{Artime2018,Artime2019, Peralta2020a, Peralta2020b,Abella2022, Baron2022, Abella2023}. As in the presence of aging the consensus state occupies a larger region of parameter space, one gets the somewhat paradoxical result that in such model an increase in the resistance to change state helps to reach a consensus situation. The question then arises about whether such tendency is model dependent or can occur in other opinion models.
 
Beyond the canonical voter model and its variants, another important class of opinion formation models is that based on kinetic exchanges~\cite{ Lallouache2010,Sen2011,Biswas2011,Biswas2012,Crokidakis2012}. In these models, the influenced individual can retain part of its original opinion, and besides the two extreme polarities, that can be represented by $\pm 1$ as in the voter model, the neutral value 0, or other intermediate values, are possible. Within this class, we consider a noisy version where, additionally to the kinetic exchange, there is the possibility of randomly adopting a new opinion, independently of the contacts with other individuals~\cite{Crokidakis2014,Vieira2016}. In this case, unlike the noisy voter model, it has been reported the existence of a non-zero critical point even in the thermodynamic limit~\cite{Crokidakis2014}. Then, we address the effects of aging in such case. We will see that the critical point can be shifted in the direction of disfavoring consensus, although the opposite can also occur for a given range of model parameters.

The outline of the paper is as follows. In Section~\ref{sec:model}, we describe the dynamics governed by stochastic rules of update of the opinion states and opinion age of the agents. 
In Section~\ref{sec:MF}, we present a complete description of the system in the mean-field limit giving the rate equations of the process and find the phase diagram, for different forms of the age-dependent rates, and in terms of the parameters of the model. The analytical results are supported by those coming from numerical simulations in complete graphs for particular forms of age-dependency, compared to theoretical calculations, presented in Section~\ref{sec:results}. Conclusions and final considerations are discussed in Section~\ref{sec:final}.

\section{Model}
\label{sec:model}

The system consists of $N$ agents, or voters, connected by links. Throughout this study, we exclusively consider the all-to-all connected topology, or complete graph, wherein each agent is linked to every other one. Each agent $i=1,\dots,N$ holds an opinion, or state variable $s_i \in \{ -1, 0, +1\}$. The interpretation of this variable may vary depending on the context, such as the political alignment of voters along the left-right spectrum. However, this paper does not delve into the precise explication of its meaning. In addition, the age $\tau_i$ of agent $i$ represents the number of successive events in which the agent has been selected but remained in its current state. As initial condition, the state of each agent is randomly selected among the three possible states and the age of each agent is set equal to 0. 

We consider the following updating rules based on kinetic exchanges in pairwise interactions and independence (noise)~\cite{Crokidakis2014}, with the introduction of aging effects: At each iteration, an agent $i$ is randomly selected. This agent can either randomly change its state, with probability $a$, or interact with another agent $j$, randomly selected from the set of all its neighbors, with probability $1-a$. In the former case, a new state $s_i$ is randomly selected from the three possible values $s_i=-1,0,1$, irrespective of its previous value. In the latter case when a contact with agent $j$ has been established, the new state of agent $i$ depends on the following scenarios:
\begin{itemize}
 \item With probability $q(\tau_i)$, where $q(\tau_i) $ models the persistence or reaction to change of the individual $i$ as a function of the age $\tau_i$, agent $i$ changes its state according to the kinetic rule
\begin{eqnarray}\label{eq:kinetic_rule}
 s_i & \rightarrow {\rm sgn}[s_i+s_j],
\end{eqnarray}
where ${\rm sgn}[s]$ is the sign function.
\item With complementary probability $1-q(\tau_i)$, nothing happens.
\end{itemize}
Independently of the update mechanism actually used by agent $i$ (random change or pairwise interaction) its age $\tau_i$ changes in the following way:
\begin{itemize}
 \item If agent $i$ changes its state, its age is reset, i.e., $\tau_i\to0$.
 \item If agent $i$ does not change its state, its age is incremented in one unit, i.e., $\tau_i\to\tau_i+1$. 
\end{itemize}

As on average an agent is updated once every $N$ steps, the age $\tau$ is measured in units of Monte Carlo steps (MCS), where $1$ MCS = $N$ individual update attempts. 

In the absence of inertia or aging, it is $q_i=1$ for all agents $i=1,\cdots,N$. Otherwise $q(\tau_i)$ is a non constant function of the age $\tau_i$. We will consider the following functional forms for the age-decaying probabilities: algebraic, $q(\tau)=1/(1+\tau/\tau^*)$, and exponential decay, $q(\tau)=\exp(-\tau/\tau^*)$. Note that in both cases the aging-less model is recovered for $\tau^*\to\infty$. For completeness and comparison with previous results, we also consider the so-called {\slshape anti-aging} scenario, where the probability $q(\tau)$ is now an increasing function of age~\cite{Peralta2020a}. This situation reflects those cases in which agents get more prone to change state the longer they have been holding the current state. An specific functional form for anti-aging is $q(\tau)=(q_0+\tau/\tau^*)/(1+\tau/\tau^*)$, with $0<q_0<1$, which recovers the aging-less model when $q_0\to 1$, for any $\tau^*$. All these forms have been previously considered in the context of the modified voter model~\cite{Baron2022}. 

\section{Mean-field description}
\label{sec:MF}
Let $x^s_\tau$ be the fraction of agents in state $s=-1,0,+1$ and age $\tau$ relative to the total population $N$. Hence, the total fraction of agents in state $s$ is $x^s=\sum_{\tau=0}^{\infty}x^s_\tau$.
The mean-field rate equations of $x^s_\tau$ can be written, for $\tau\geq1$, as (see Appendix~\ref{app_xtau} for details of their derivation)
\begin{equation}\label{eq:ratesi1}
 \begin{split}
 \frac{dx_\tau^+}{dt}=&a \Bigl(\frac{1}{3}x_{\tau-1}^+-x_\tau^+\Bigr)+(1-a)\times\\
 &\left(x^+_{\tau-1}(1-q_{\tau-1})+ x^+_{\tau-1} q_{\tau-1}(x^++x^0)-x^+_\tau\right),\\
 \frac{dx_\tau^0}{dt}=&a \left(\frac{1}{3}x_{\tau-1}^0-x_\tau^0\right)+(1-a)\times\\
 &\left(x^0_{\tau-1}(1-q_{\tau-1})+ x^0_{\tau-1} q_{\tau-1} x^0-x^0_\tau\right), \\
 \frac{dx_\tau^-}{dt}=&a \left(\frac{1}{3}x_{\tau-1}^--x_\tau^-\right)+(1-a)\times\\
 &\left(x^-_{\tau-1}(1-q_{\tau-1})+ x^-_{\tau-1} q_{\tau-1}(x^-+x^0)-x^-_\tau\right), 
 \end{split}
\end{equation}
where for brevity in the notation we write $q_\tau$ for $q(\tau)$. For the particular case $\tau=0$, which corresponds to individuals that have just changed state, we have a different set of equations, namely,
\begin{equation} \label{eq:ratesi0}
 \begin{split}
 \frac{dx_0^+}{dt}&=a \left(\frac{1}{3}\left(1-x^+\right)-x_0^+\right)+\\
 &+(1-a)\left(-x^+_0+ x^+ y^0 \right),\\
 \frac{dx_0^0}{dt}&=a \left(\frac{1}{3}\left(1-x^0\right)-x_0^0\right)+\\
 &+(1-a)\left(-x^0_0+x^+ y^- +x^- y^+ \right),\\
 \frac{dx_0^-}{dt}&=a \left(\frac{1}{3}\left(1-x^-\right)-x_0^-\right)+\\
 &+(1-a)\left(-x^-_0+x^-y^0 \right),
 \end{split}
\end{equation}
where we have defined 
\begin{equation} \label{eq:ys}
 y^s \equiv \sum_{\tau=0}^{\infty} q_\tau x^s_\tau,\quad s=-1,0,+1.
\end{equation}
 
Adding Eqs.~(\ref{eq:ratesi1}) and (\ref{eq:ratesi0}) over all values of $\tau$, we obtain the equations for the density of each opinion, 
\begin{equation} \label{eq:rates+0-}
 \begin{split}
\frac{d x^{+}}{dt} &=a\left(\frac{1}{3}-x^+\right) + (1-a)(x^+y^0 -x^-y^+), \\ 
\frac{d x^{0}}{dt} &=a\left(\frac{1}{3}-x^0\right) + \\
&\quad (1-a)( x^+ y^- + x^- y^+ -(x^-+x^+)y^0 ), \\
\frac{d x^{-}}{dt} &=a\left(\frac{1}{3}-x^-\right) + (1-a)(x^- y^0 -x^+ y^-).
\end{split}
\end{equation}
Note that, as $\frac{d}{dt}(x^++x^-+x^0)=0$, the normalization condition $x^++x^-+x^0=1$ is satisfied at all times, as it should be. 

Our aim now is to derive a closed set of evolution equations for the global variables $x^+$, $x^0$ and $x^-$. To this end, we need to express the variables $y^+,y^-,y^0$ appearing in Eqs.~\eqref{eq:rates+0-} in terms of $x^+$ and $x^-$. 

The first step is the use of an adiabatic approximation whereby we set to zero the left-hand side of Eqs.~(\ref{eq:ratesi1}). This leads to recursion relations for $x_\tau^s$ in terms of $x_{\tau-1}^s$ for $\tau\ge 1$. The solutions of those recursive relations are
\begin{equation} \label{eq:ratesi}
\begin{split}
x_\tau^+&=x_0^+ F_\tau(x^-),\\
x_\tau^0&=x_0^0 F_\tau(1-x^0),\\
x_\tau^-&=x_0^- F_\tau(x^+),
\end{split}
\end{equation}
where we have introduced
\begin{equation} \label{eq:Ftau}
F_\tau(x)\equiv \prod_{k=0}^{\tau-1}\gamma(q_k\, x,a),\quad \tau\ge 1,
\end{equation}
with
\begin{equation} 
\label{eq:gamma_def}
\gamma(z,a)\equiv\frac{a}{3}+(1-a)(1-z).
\end{equation}
For $\tau=0$, to ensure consistence in the notation of Eqs.~\eqref{eq:ratesi}, we define $F_0(x)=1$. 
Summing the densities $x_\tau^+$, $x_\tau^0$, $x_\tau^-$, over all values of $\tau=0,1,\dots$, we obtain
\begin{equation} \label{eq:xxx}
\begin{split}
x^+ &= x_0^+\sum_{\tau=0}^{\infty}F_\tau(x^-),\\ 
x^0 &= x_0^0\sum_{\tau=0}^{\infty}F_\tau(1-x^0),\\
x^- &= x_0^-\sum_{\tau=0}^{\infty}F_\tau(x^+),
\end{split}
\end{equation}
which, combined with Eqs.~\eqref{eq:ratesi}, allows one to express the aging-dependent densities $x_\tau^+$, $x_\tau^0$, $x_\tau^-$, in terms of the global ones, $x^+$, $x^0$ and $x^-$, as 
\begin{equation} \label{eq:xxx_tau}
 \begin{split}
 x_\tau^+&=x^+\frac{F_\tau(x^-)}{\sum_{\tau=0}^\infty F_\tau(x^-)},\\
 x_\tau^0&=x^0\frac{F_\tau(1-x^0)}{\sum_{\tau=0}^\infty F_\tau(1-x^0)},\\
 x_\tau^-&=x^-\frac{F_\tau(x^+)}{\sum_{\tau=0}^\infty F_\tau(x^+)}.
 \end{split}
\end{equation}
Replacing these expressions in the definitions of Eqs.~\eqref{eq:ys}, we obtain
\begin{equation} \label{eq:yyy}
\begin{split}
y^+ &=x^+ \Phi(x^-), \\ 
y^0 &= x^0 \Phi(1-x^0), \\ 
y^- &= x^- \Phi(x^+), 
\end{split}
\end{equation}
where we have introduced the function
\begin{equation} \label{eq:Phi}
\Phi(x) \equiv 
\frac{\sum_{\tau=0}^\infty q_\tau F_\tau(x)}{\sum_{\tau=0}^\infty F_\tau(x)}.
\end{equation}
Let us note here, for consistency, that in the aging-less case, $q_\tau=1$, it is $\Phi(x)= 1$ and, hence, $y^s=x^s$, for $s=-1,0,+1$, in agreement with the definition in Eq.~(\ref{eq:ys}). 

Replacement of Eqs.~\eqref{eq:yyy} in Eqs.~\eqref{eq:rates+0-}, leads to the desired closed system of differential equations for $x^+$, $x^0$ and $x^-$. As $x^0=1-~x^--~x^+$ due to the normalization condition, this can be further simplified to a closed system of two differential equations for $x^-$ and $x^+$, namely, 
\begin{eqnarray}\label{dxpdt}
\frac{dx^+}{dt}&=&F(x^+,x^-),\\ \label{dxmdt}
\frac{dx^-}{dt}&=&F(x^-,x^+),
\end{eqnarray} 
where
\begin{eqnarray}\nonumber 
F(z,w)&=&(1-a)\left[z(1-z-w)\Phi(z+w)-z\, w\,\Phi(w)\right] \\ \label{eq:Fxy} 
&&+a\left(\frac13-z\right).
\end{eqnarray} 
Equations~(\ref{dxpdt}-\ref{eq:Fxy}) are the basis of our subsequent theoretical analysis. They depend on the important function $\Phi(x)$, which is determined solely by the noise intensity $a$ and the functional form of the aging function $q_\tau$.

It is important to notice that the structure of the dynamical Eqs.~(\ref{dxpdt}-\ref{eq:Fxy}) allows for the symmetric (S) steady-state solution $x^{+}_{\text{\tiny S}}=x^{-}_{\text{\tiny S}}\equiv x^{\pm}_{\text{\tiny S}}$, satisfying the equation $F(x^{\pm}_{\text{ \tiny S}},x^{\pm}_{\text{\tiny S}})=0$. Other, asymmetric (A) solutions $x^{+}_{\text{\tiny A}}$, $x^{-}_{\text{\tiny A}}$ might be possible for specific values of the system parameters. The stability of the different steady-state solutions is determined by the eigenvalues of the Jacobian matrix 
\begin{equation}\label{jacobian}
J=\left.\begin{pmatrix}\partial_z F(z,w)& \partial_w F(z,w)\\
\partial_z F(w,z)&\partial_w F(w,z)
\end{pmatrix}\right|_{z=x^+,w=x^-},
\end{equation}
evaluated at the different fixed points. 

Furthermore, note that if $(x^+,x^-)$ is a steady-state solution of the dynamical equations, it turns out that an exchange of the values of $x^+$ and $x^-$ leads also to a steady-state solution with the same stability. In the cases of interest (with the noticeable exception of the aging-less situation considered next), the steady-state solutions of Eqs.~(\ref{dxpdt})-(\ref{eq:Fxy}) and their stability have to be found numerically. By normalization, the value of the density of agents in the zero state is $x^0_\text{\tiny S}=1-2x^\pm_\text{\tiny S}$ and $x^0_\text{\tiny A}=1-x^+_\text{\tiny A}-x^-_\text{\tiny A}$ in the symmetric and asymmetric cases, respectively.

In the next section we will obtain the function $\Phi(x)$ for specific functional forms of $q_\tau$. This will be used to determine the steady-state solutions $x^s,\,s=-1,0,+1$, resulting from setting the left-hand side of Eqs.~(\ref{dxpdt}-\ref{dxmdt}) to zero, as well as their stability. 

\section{Results for particular forms of $q(\tau)$}
\label{sec:results}

\begin{figure*}[t]
\includegraphics[width=0.5\columnwidth]{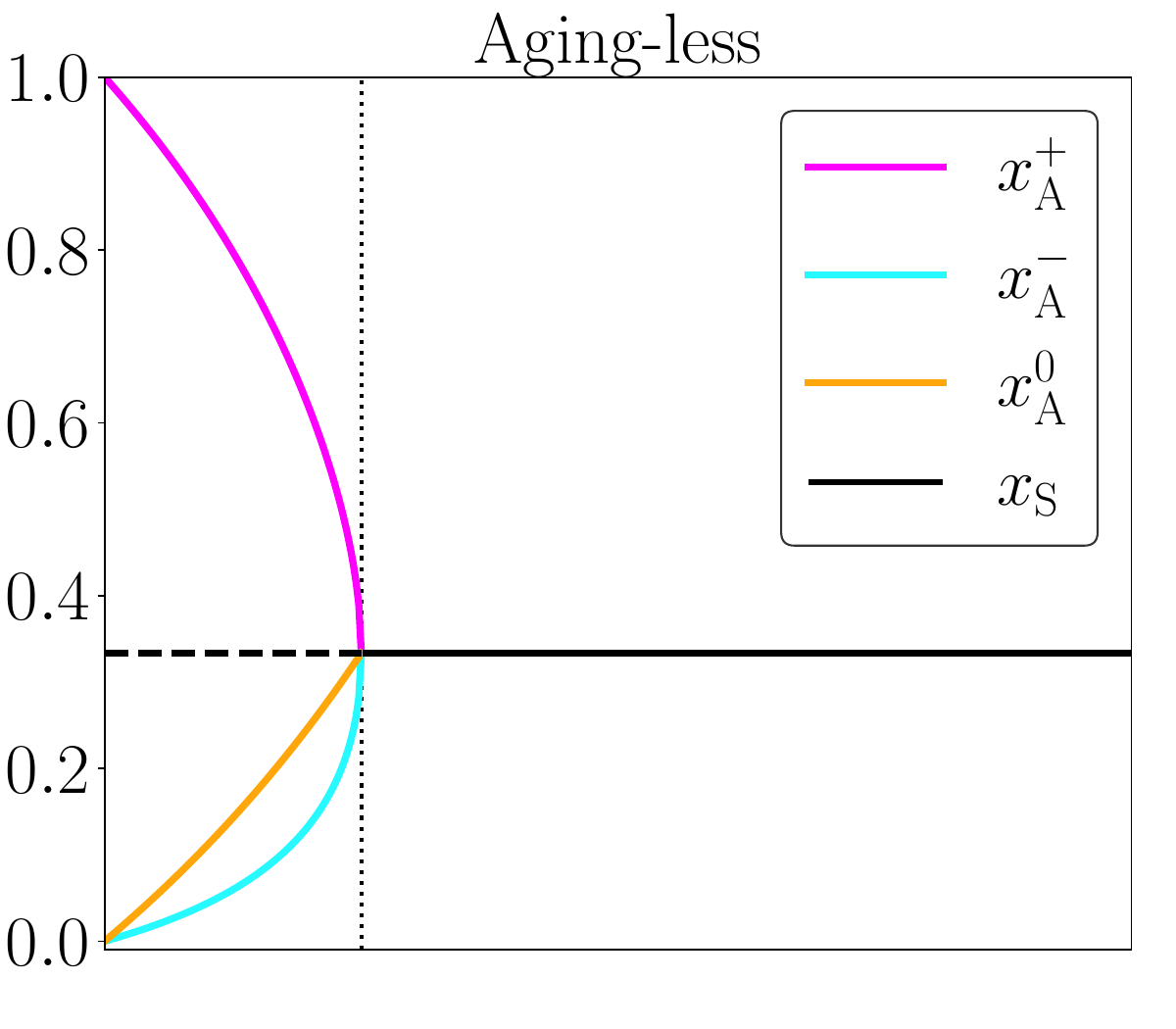}
\includegraphics[width=0.5\columnwidth]{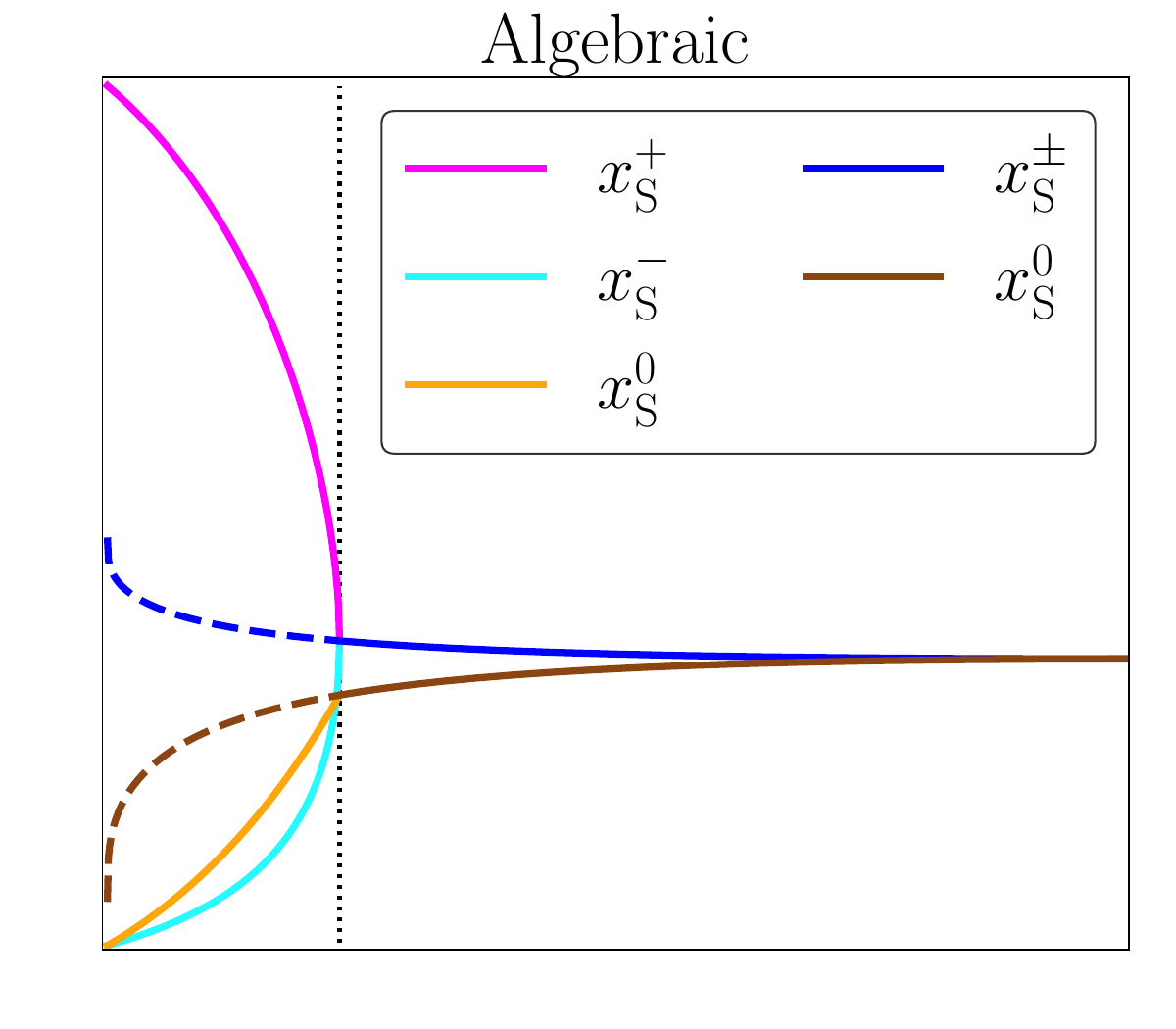}
\includegraphics[width=0.5\columnwidth]{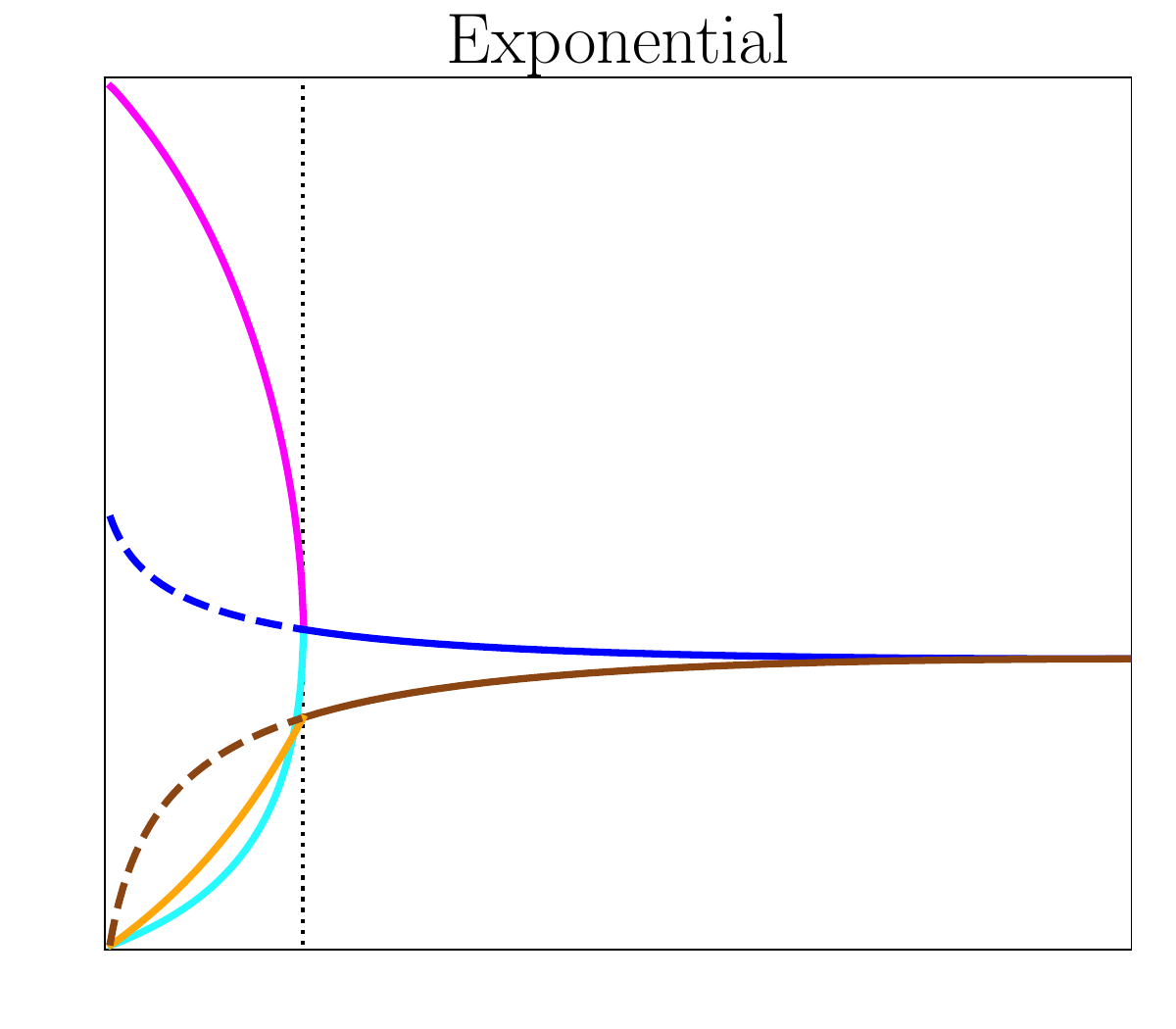}
\includegraphics[width=0.5\columnwidth]{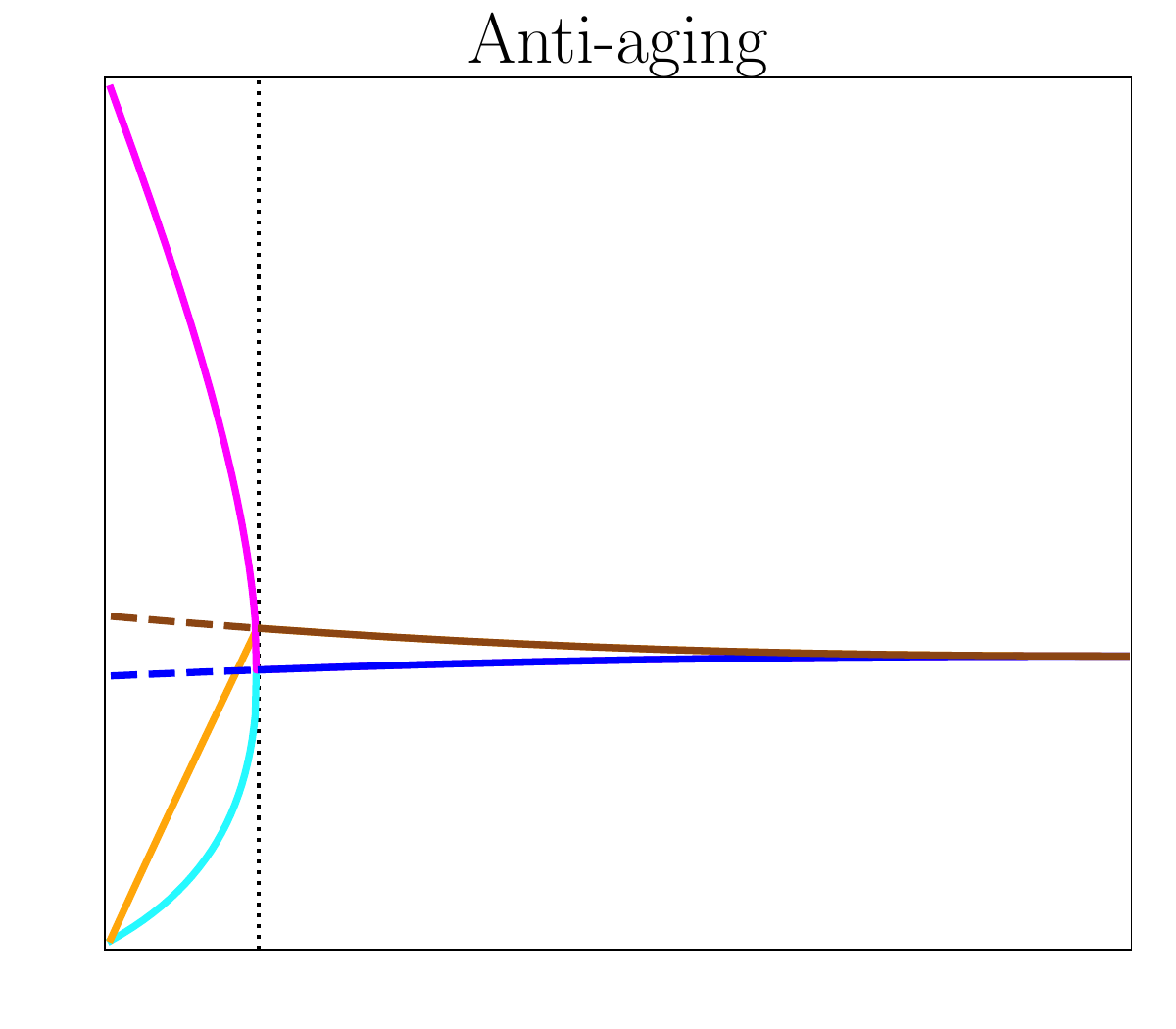}
\includegraphics[width=0.5\columnwidth]{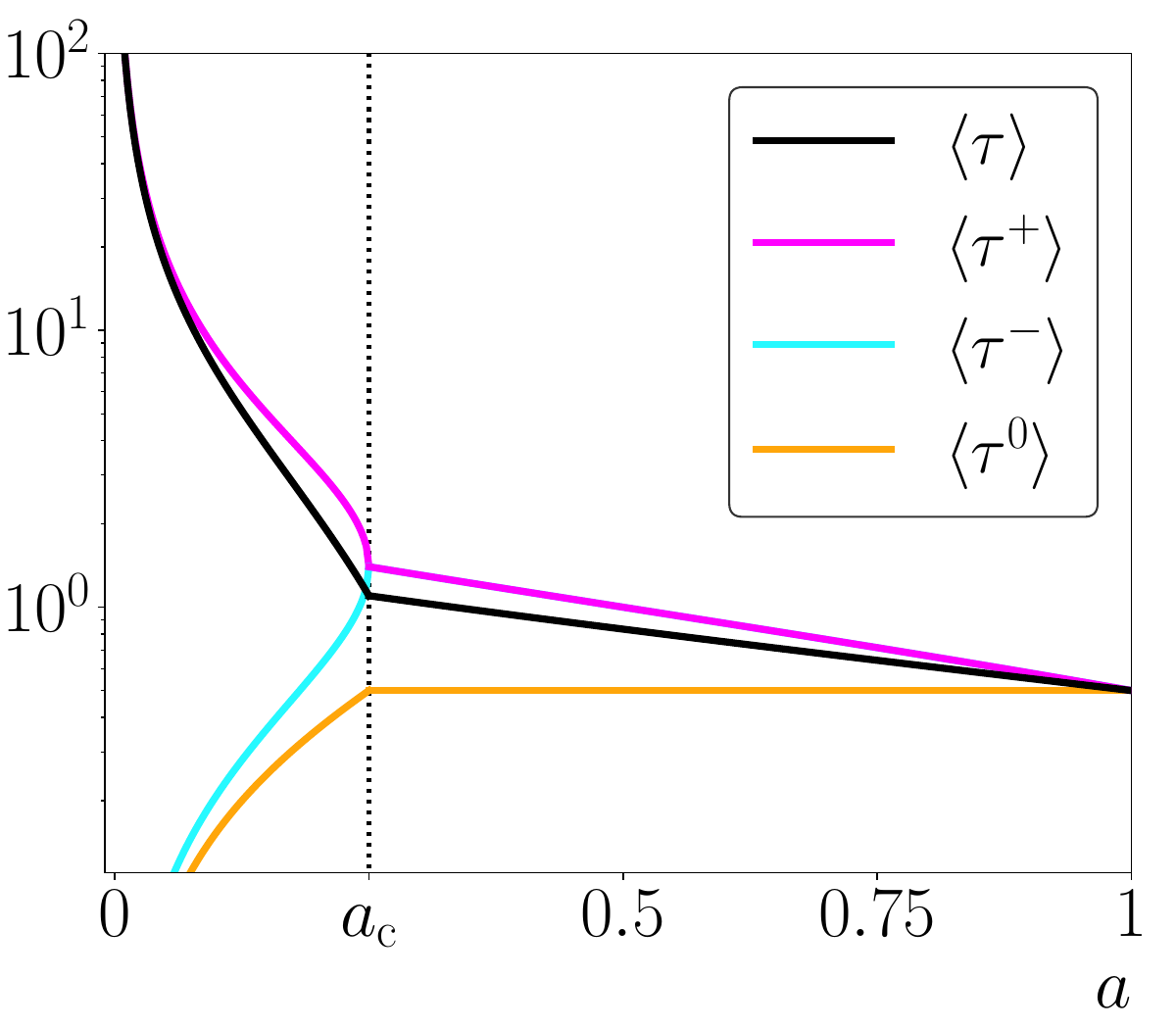}
\includegraphics[width=0.5\columnwidth]{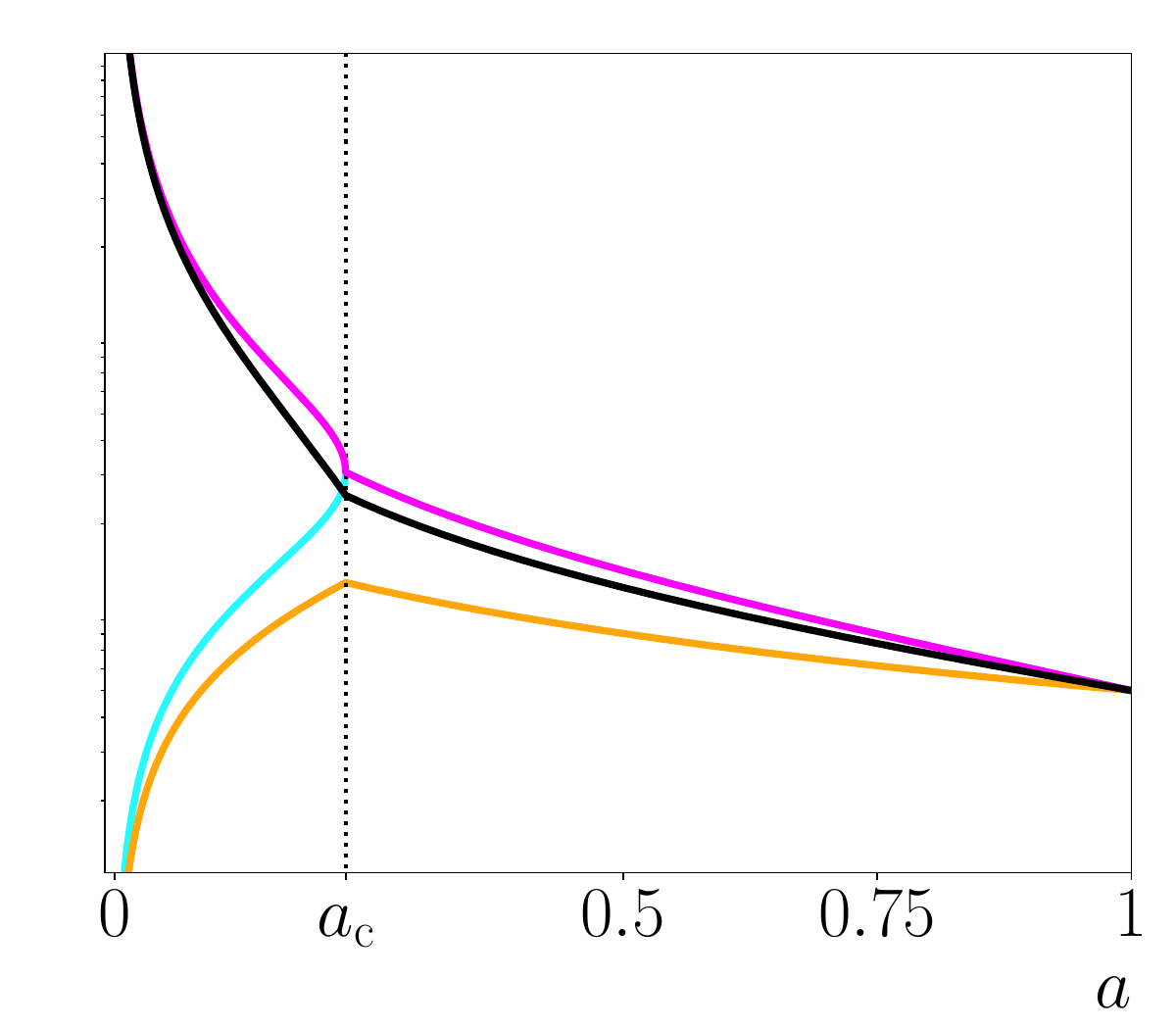}
\includegraphics[width=0.5\columnwidth]{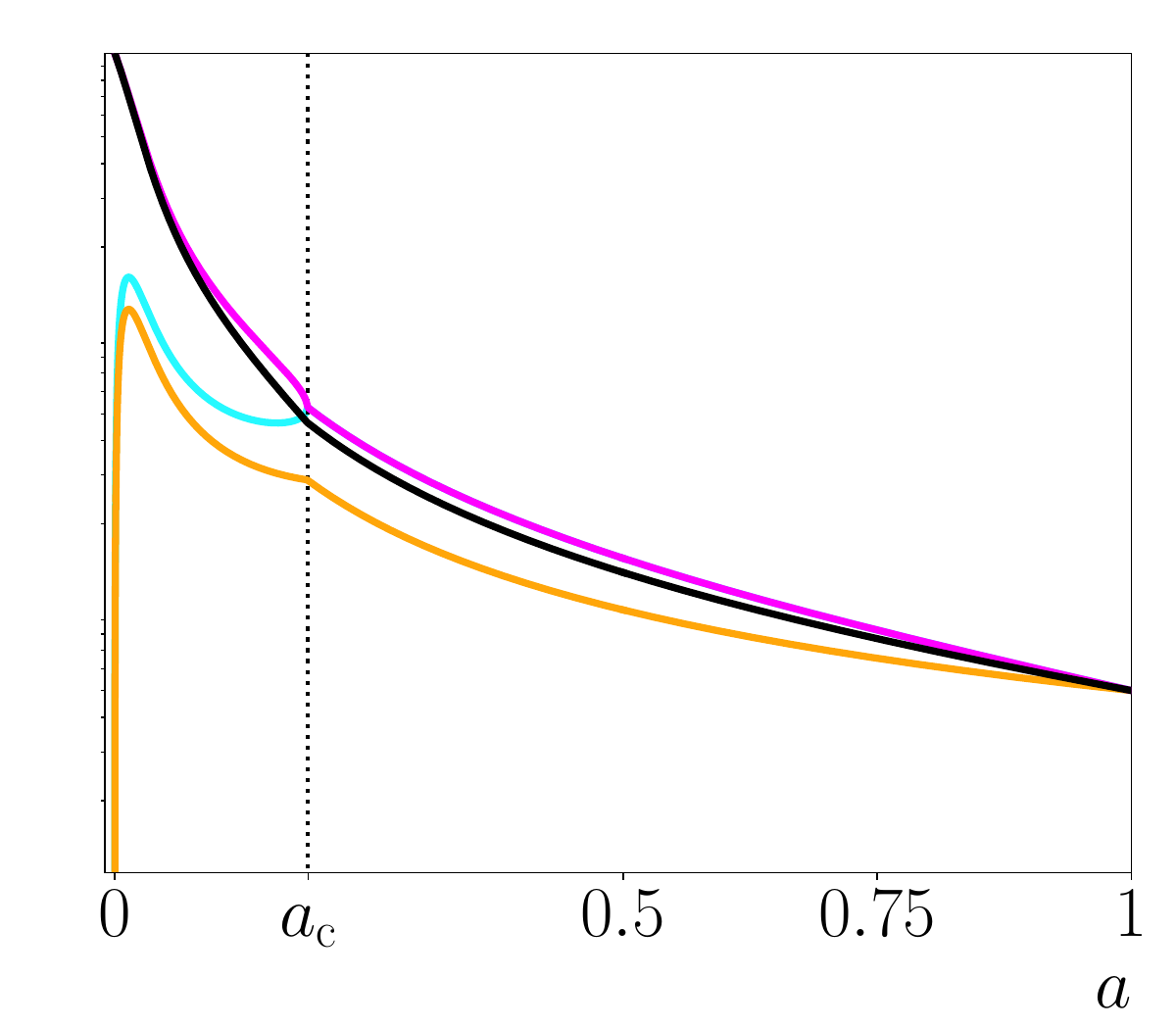}
\includegraphics[width=0.5\columnwidth]{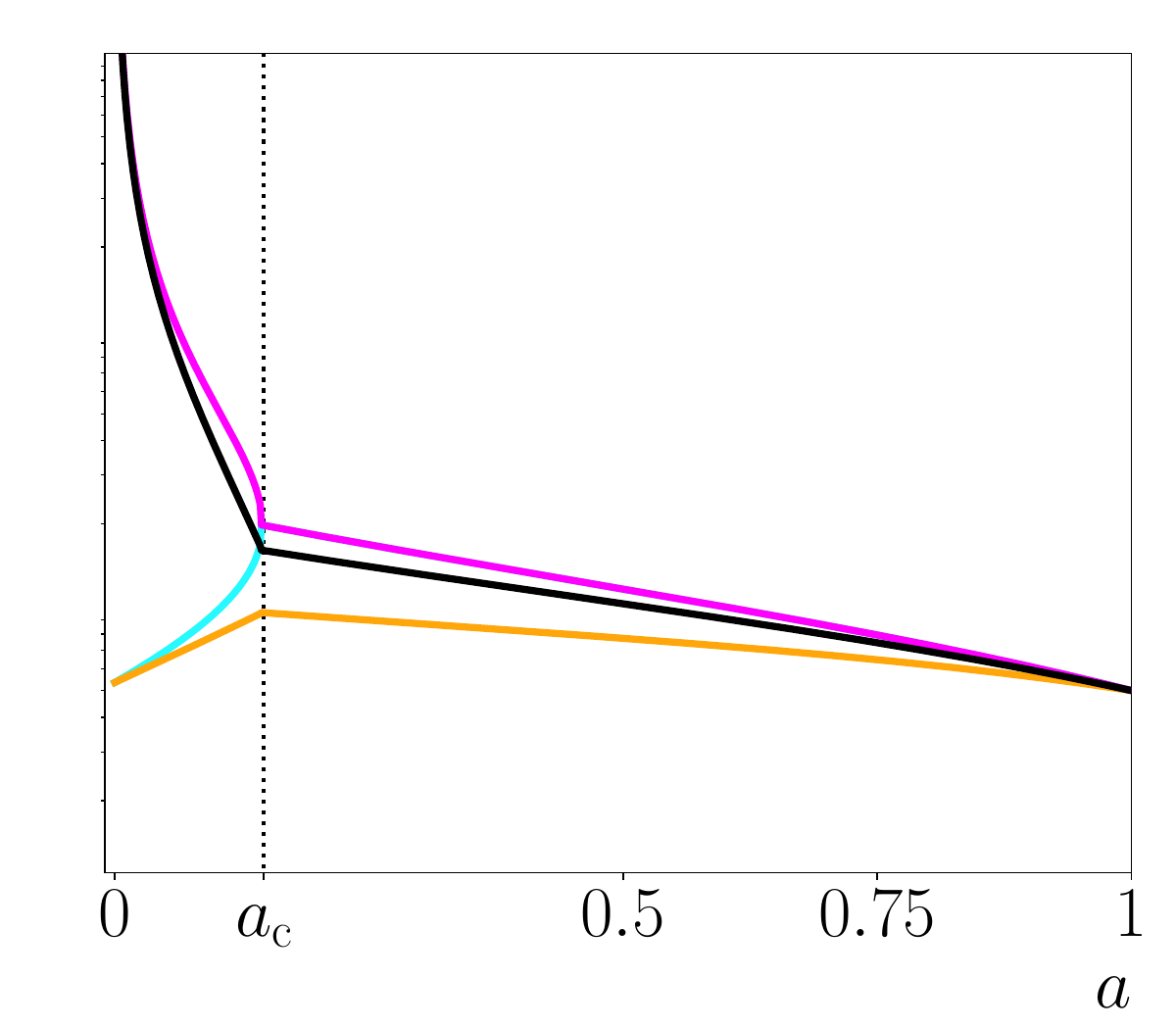}
\caption{ 
Top row: Phase diagram representing the fraction of agents $x^s$ in each state $s=-1,0,+1$ plotted as a function of the noise intensity $a$. The subscript $\text{S}$ ($\text{A}$) indicates symmetric (asymmetric) solution. In the aging-less case, the superindex of the symmetric solution has been omitted since all three fractions coincide, $x^+_\text{\tiny S}=x^-_\text{\tiny S}=x^0_\text{\tiny S}\equiv{x_\text{\tiny S}}=1/3$. The legend of the exponential and anti-aging cases is the same than that of the algebraic one. Low row: Average ages $\langle\tau^s\rangle$ for each one of the agent's states $s$, and the overall average time $\langle\tau\rangle$, vs. $a$. Same legend in all figures. Each column corresponds to a particular kernel $q(\tau)$, indicated on the top of the figure: $q_\tau=1$ for all $\tau$ (aging insensitive), $q_\tau=1/(1+\tau/\tau^*)$ (algebraic), $q_\tau=\text{exp}(-\tau/\tau^*)$ (exponential), and $q_\tau=(q_0+ \tau/\tau^*)/(1+ \tau/\tau^*)$ (anti-aging). In all cases we have assumed that a majority of agents are in state $+1$ in the disordered state. Vertical dashed lines indicate the critical value $a_\text{c}$, which is determined with Eq.~\eqref{eq:noaging-mag} in the aging-less case, $a_\text{c}=0.25$ and numerically in the other cases obtaining: $a_\text{c}=0.22181686689\dots$, $a_\text{c}=0.18924447367\dots$, $a_\text{c}=0.14667099357\dots$, respectively from left to right. Parameter values are: $\tau^*=2.0$ (algebraic and exponential) and $\tau^*=q_0=0.1$ (anti-aging). 
} 
\label{fig:theo}
\end{figure*}

\subsection{Aging-insensitive updating probabilities}\label{sec:aging-less}

Let us discuss first the case in which the probability $q_\tau$ is independent on the persistence time $\tau$, with $q_\tau=1$ for all $\tau$. As explained before, this leads to $\Phi(x)=1$ and the steady state solution of Eqs.~(\ref{dxpdt}-\ref{eq:Fxy}) has the following three different possibilities. 

The symmetric solution (I):
\begin{eqnarray} \label{eq:S}
x^+_\text{\tiny S}=x^-_\text{\tiny S}=x^0_\text{\tiny S}\equiv{x_\text{\tiny S}}=1/3,
\end{eqnarray}
\indent
The asymmetric solution (II): 
\begin{eqnarray} \label{eq:A}
x^0_\text{\tiny A}=\frac{a}{1-a}, \;\;
 x^{\pm}_\text{\tiny A}=\frac{3-6a\pm \sqrt{3(1-4a)(3-4a)}}{6 (1-a)}.
%%%\label{eq:solutions_agingless}\\ 
\end{eqnarray}
Solution (III) is the same as solution (II) but exchanging the values of $x^+_\text{\tiny A}$ and $x^-_\text{\tiny A}$. Note that solutions (II) and (III) only exist for $a\le~1/4$. The linear stability analysis shows that solution (I) is the stable one for $a>~1/4$ while solutions (II) and (III) are stable whenever they exist, for $a<1/4$. We plot the different solutions in Fig.~\ref{fig:theo} (first column, upper row), 
where colored lines correspond to solution (II) while the darker (black) line displays solution (I), furthermore solid lines indicate that the corresponding solution is stable, while a dashed line indicates an unstable solution. Note that, for solution (II) it is $x^+_\text{\tiny A}\ge x^0_\text{\tiny A}\ge x^-_\text{\tiny A}$, while for solution (III) the order is reversed, $x^-_\text{\tiny A}\ge x^0_\text{\tiny A}\ge x^+_\text{\tiny A}$. It is clear from this figure that the system undergoes a phase transition from a disordered phase at $a>1/4$ in which all possible states are equally present, to a symmetry breaking transition to solution (II) in which a majority of agents hold the $+1$ state, or to solution (III) with a majority in the $-1$ state. In the following, and due to the aforementioned symmetry between the two solutions, we only discuss the results corresponding to solution (II).
To characterize the phase transition we use as an order parameter the {\slshape magnetization} $m=|x^+ -x^-|$, whose stable steady-state value is greater than $0$ for $a\le a_\text{c}$ and equal to $0$ for $a\ge a_\text{c}$ with a critical value $a_\text{c}=1/4$. As a function of the noise intensity $a$, in agreement with previous results obtained by Crokidakis for the scenario where aging is absent~\cite{Crokidakis2014}, we have
\begin{equation}
 m= \begin{cases}
 \frac{\displaystyle \sqrt{(1-4a)(3-4a)}}{\displaystyle \sqrt{3} (1-a)}, &a\le a_\text{c} \\0, &a\ge a_\text{c}
 \end{cases}
 .
 \label{eq:noaging-mag}
\end{equation}

%%%%%%%%%%%%%%%%%%%%%%%%%%%%%%%%%%%%%%%%
\begin{figure*}[t!]
\includegraphics[width=0.68\columnwidth]{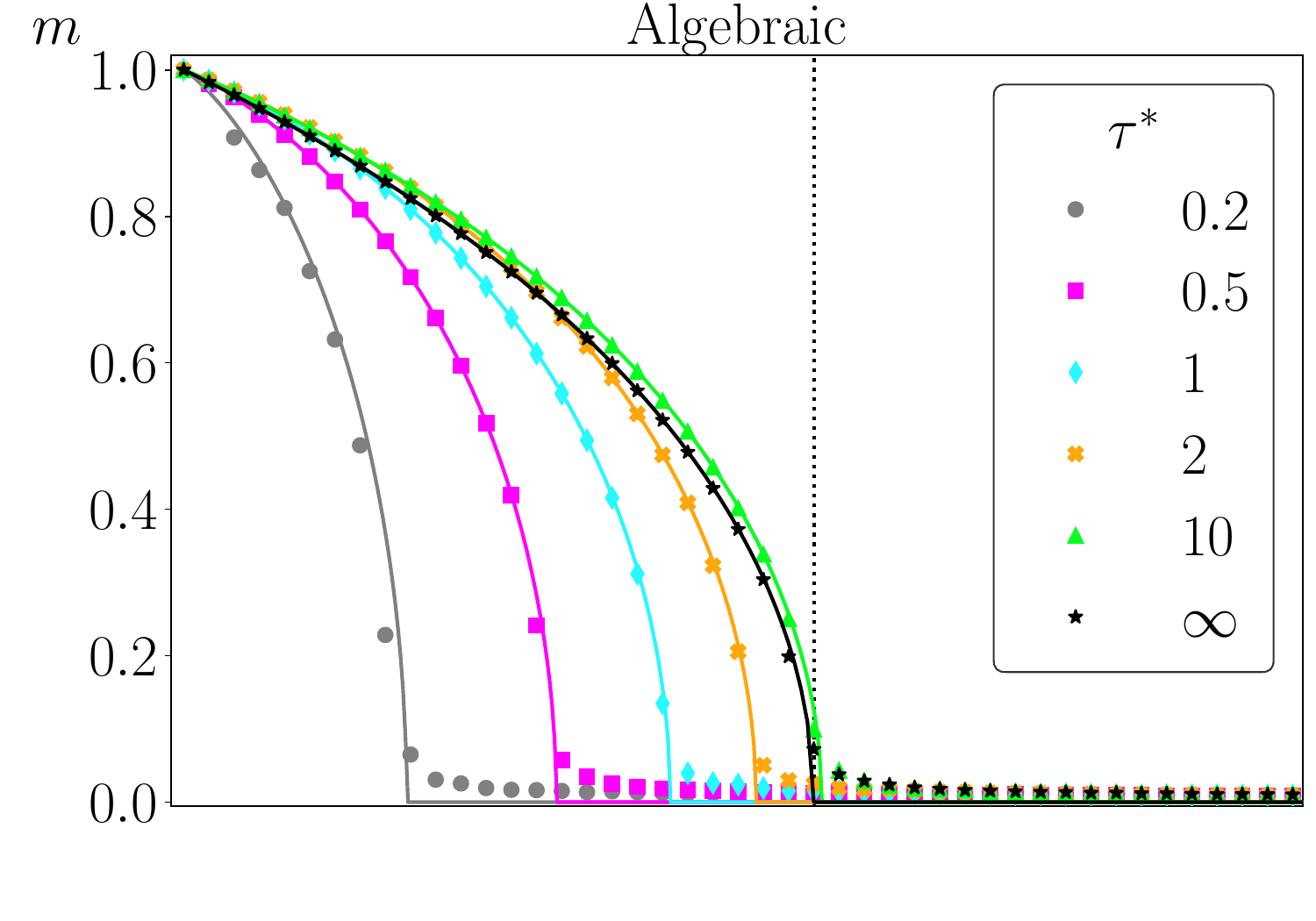}
\includegraphics[width=0.68\columnwidth]{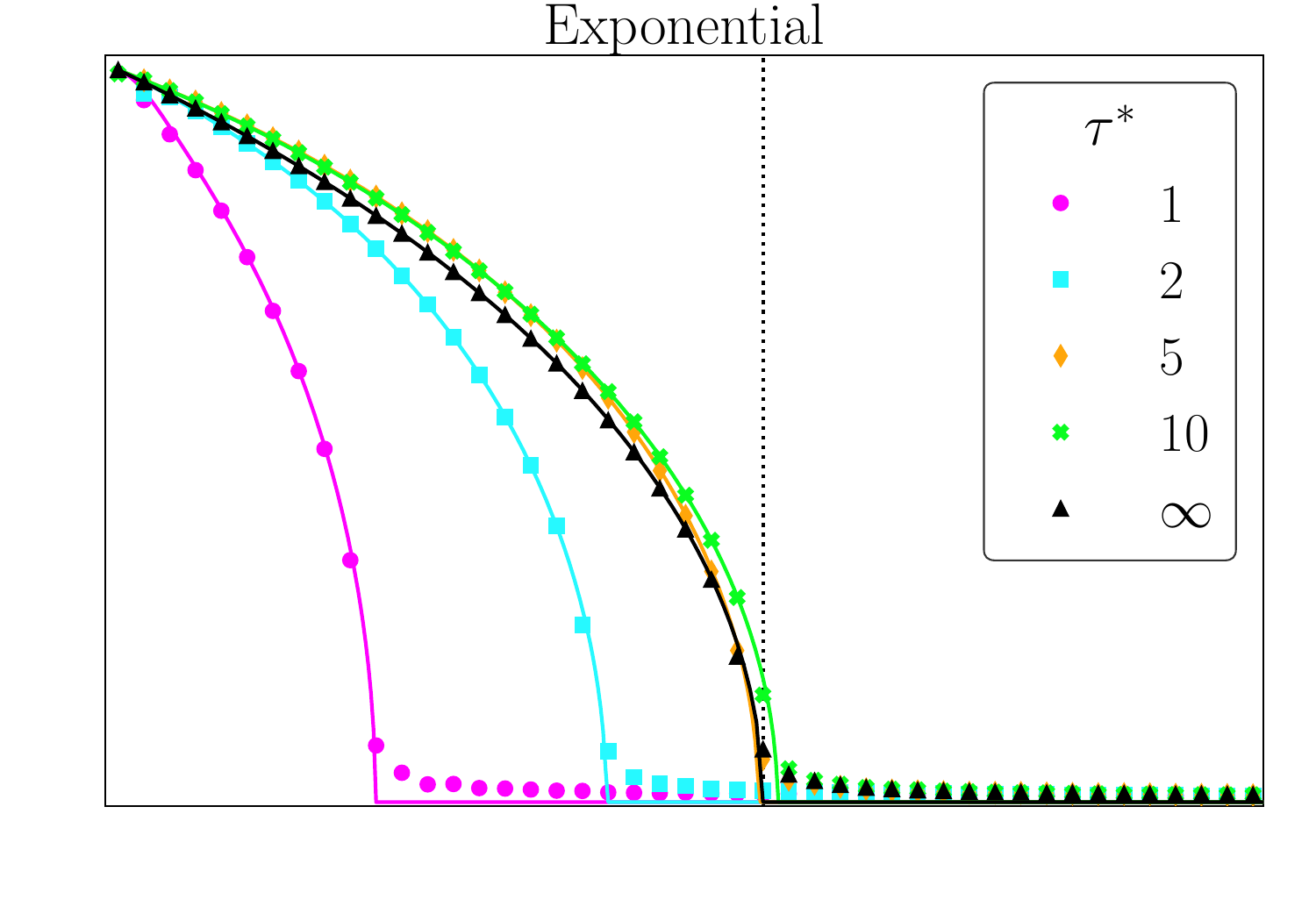}
\includegraphics[width=0.68\columnwidth]{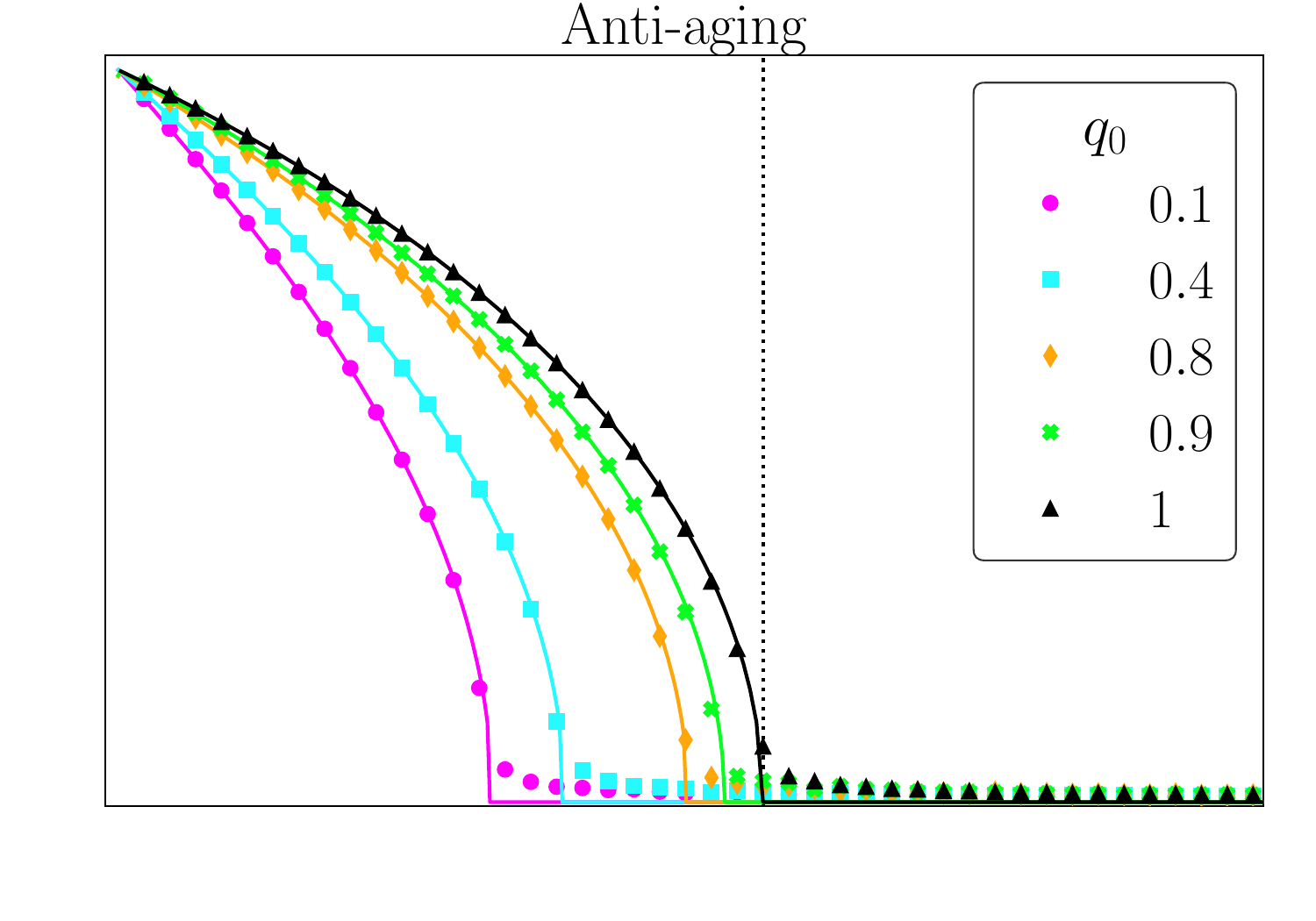}
\includegraphics[width=0.68\columnwidth]{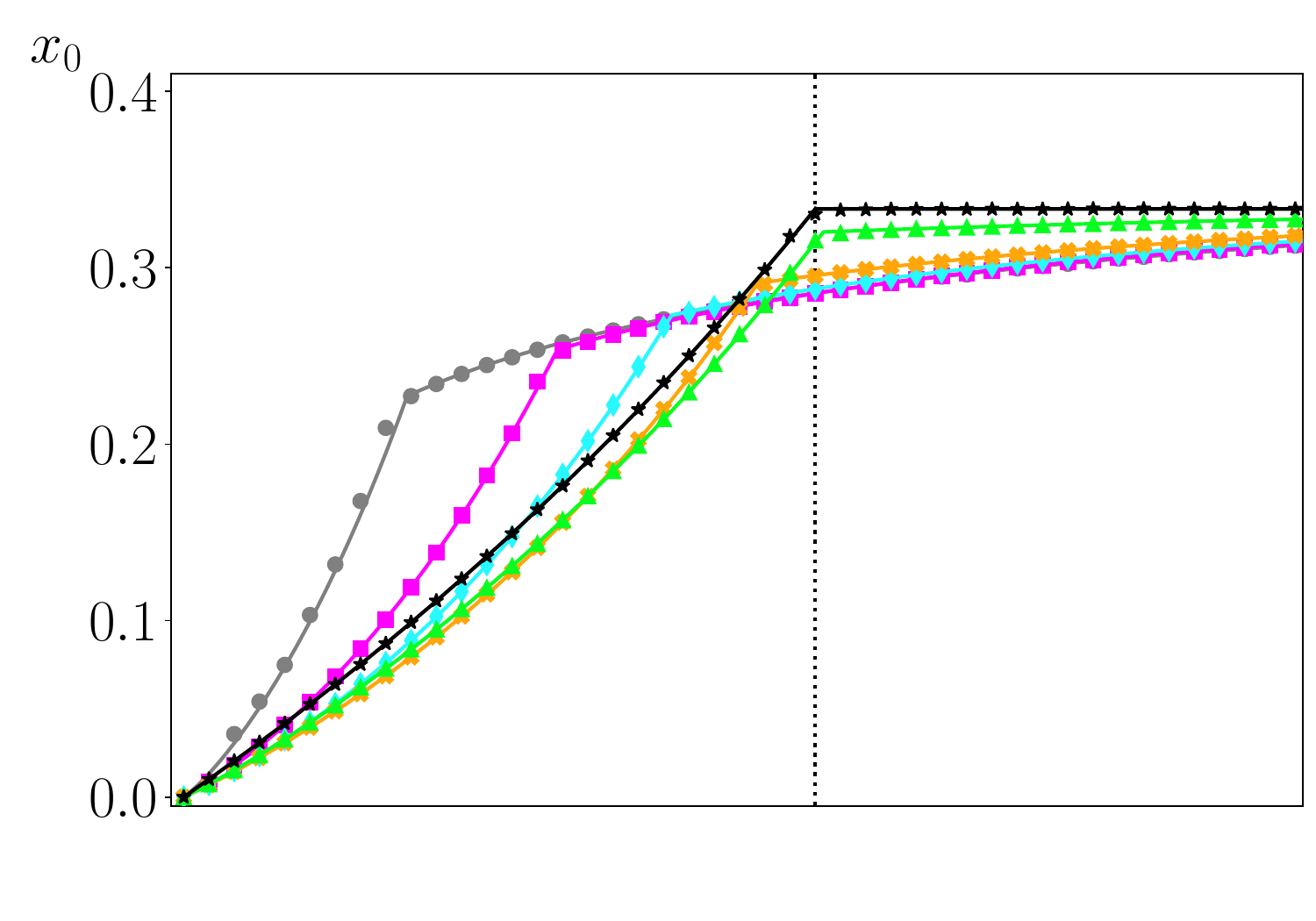} 
\includegraphics[width=0.68\columnwidth]{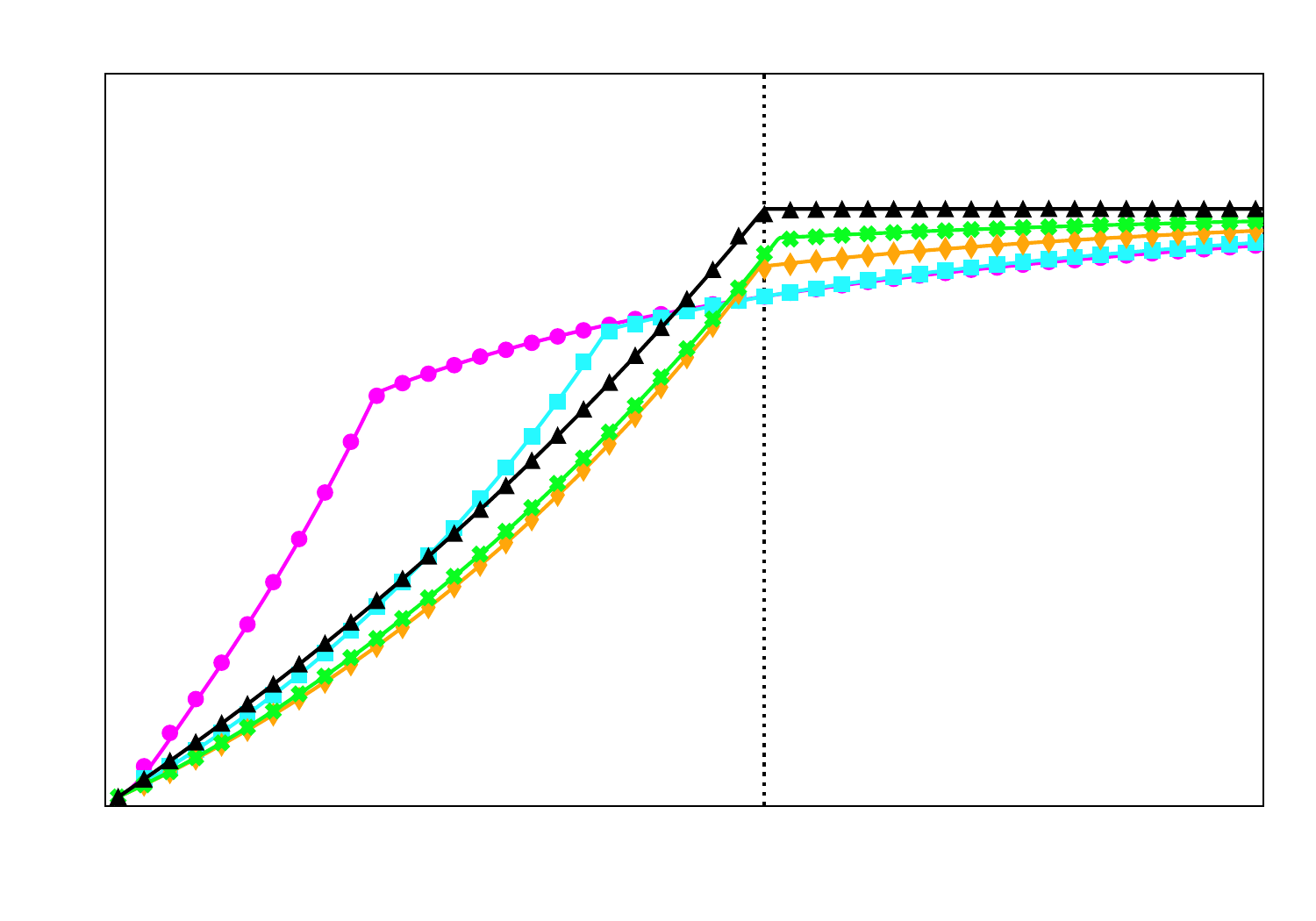}
\includegraphics[width=0.68\columnwidth]{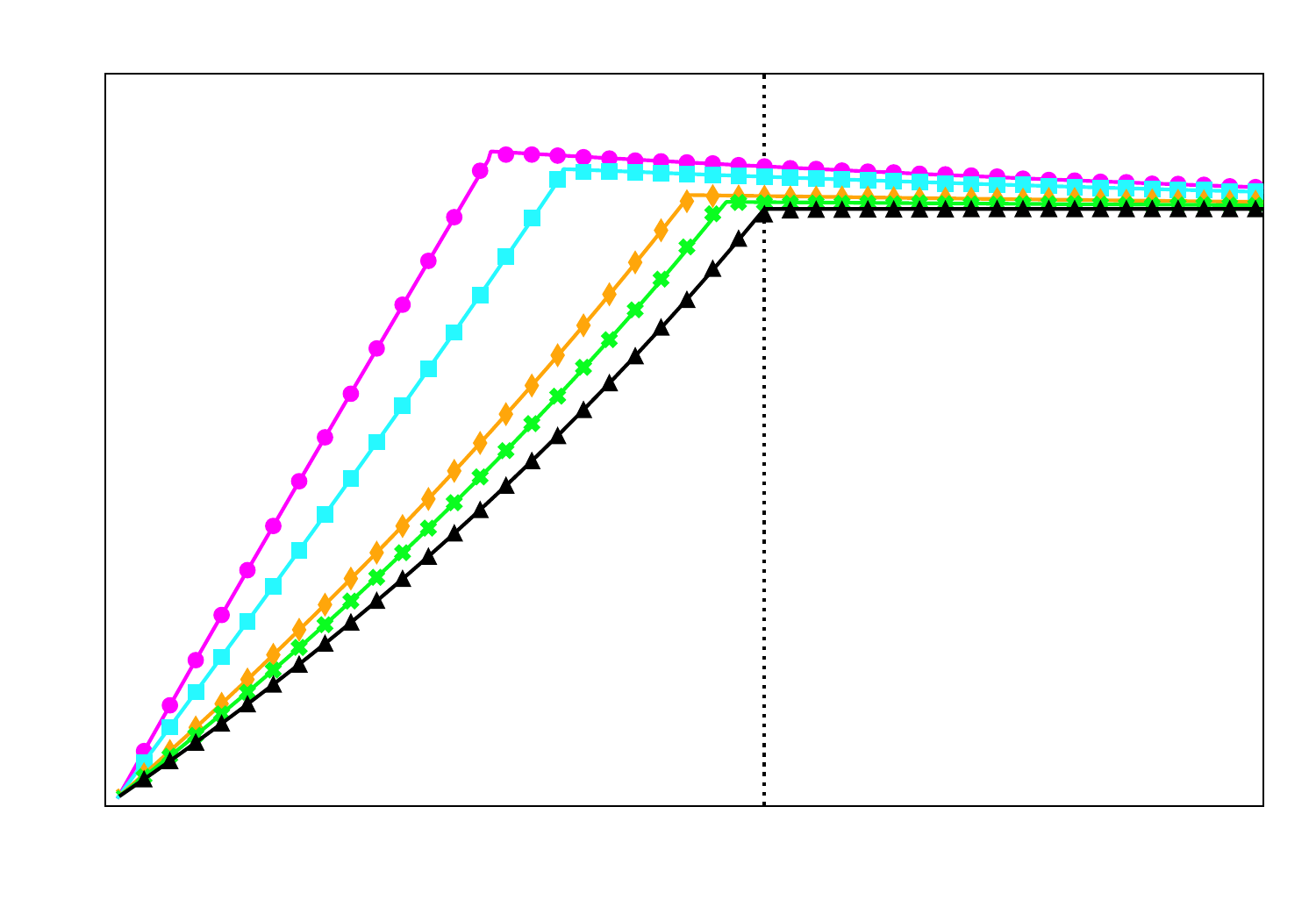}
\includegraphics[width = 0.68\columnwidth]{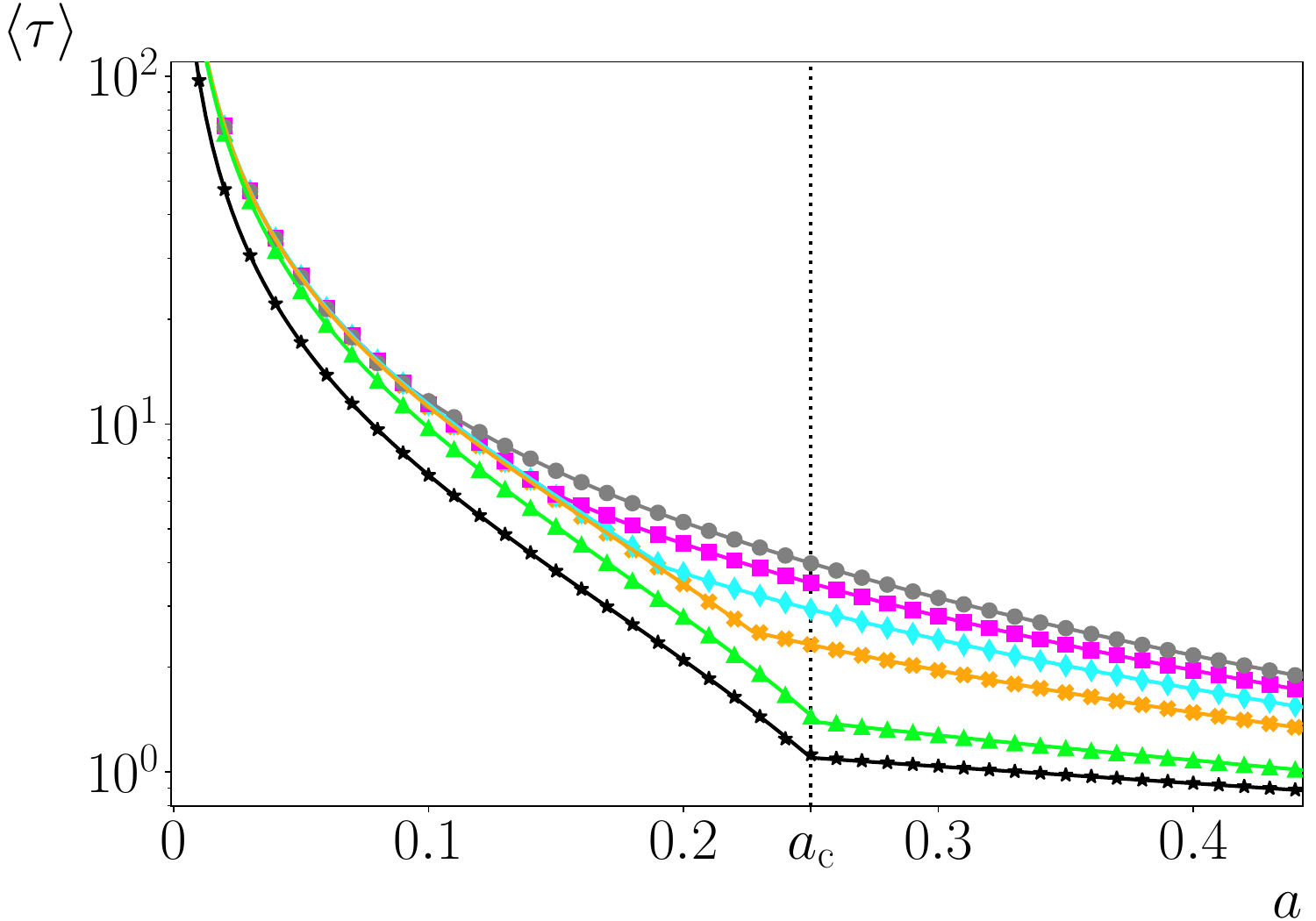}
\includegraphics[width = 0.68\columnwidth]{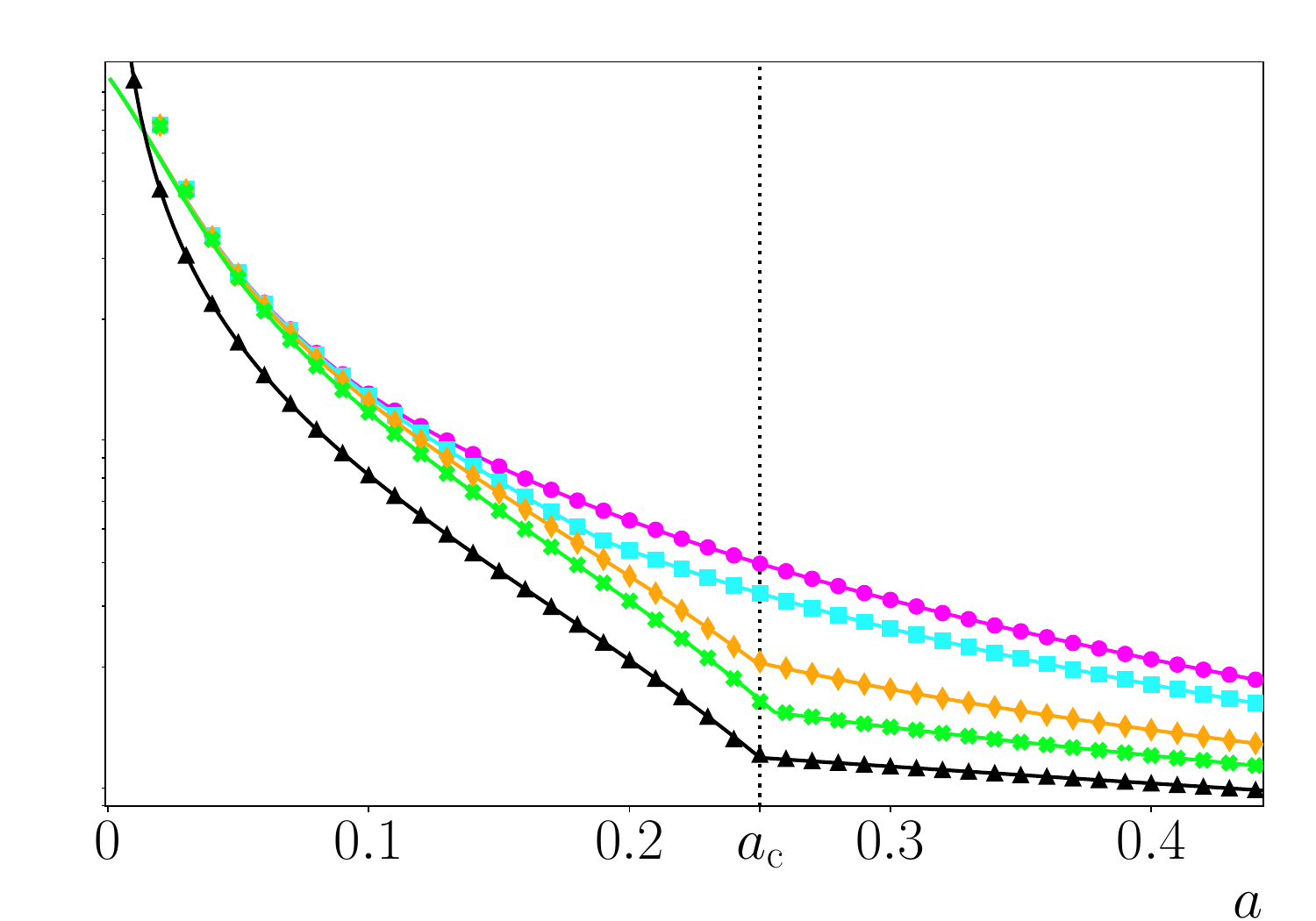}
\includegraphics[width = 0.68\columnwidth]{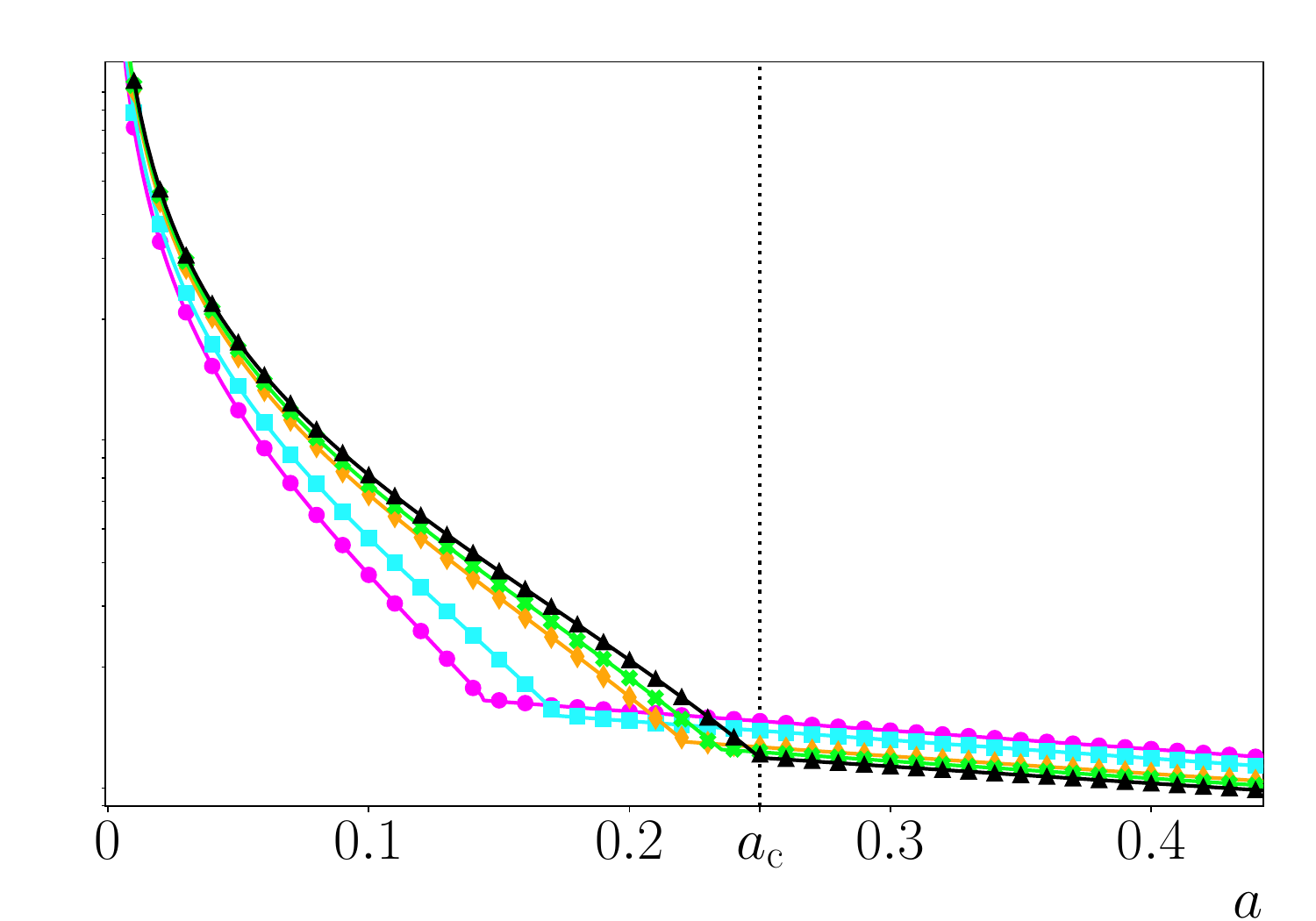}
\caption{Magnetization $m=|x^+-x^-|$ (top row), fraction $x^0$ of undecided agents (central row), and average aging time $\langle \tau \rangle$ (lower row), as a function of the noise intensity $a$ and for different values of the $\tau^*$ parameter as indicated in the legend. 
Each column corresponds to a particular kernel $q(\tau)$, indicated on the top of the figure: 
 $q_\tau=1/(1+\tau/\tau^*)$ (algebraic) and $q_\tau=\text{exp}(-\tau/\tau^*)$ (exponential), for different values of $\tau^*$ indicated in the legend, and 
$q_\tau=(q_0+ \tau/\tau^*)/(1+ \tau/\tau^*)$ (anti-aging), with $\tau^*=0.1$ and different values of $q_0$. 
The results of the mean-field theory are plotted with solid lines, while those of numerical simulations are displayed with symbols, for values of $\tau^*$. The vertical dashed line indicates the critical value $a_\text{c}$ for the aging-less case, drawn as a reference for comparison. 
The simulations have been performed using the rules of the agent-based model with $N=10^4$ agents, and averages were computed over 100 samples. 
}
\label{fig:simul}
\end{figure*}

%%%%%%%%%%%%%%%%%%%%%%%%%%%%%%%%%%%%%%%

Note that the magnetization vanishes at the critical point as $m\sim(a_\text{c}-a)^\beta$ with exponent $\beta=1/2$, as expected for the mean-field scenario.

Even though the updating probabilities are independent of age, still agents increase their age when, as a result of the dynamical rule, their state is not changed, and reset their age to $0$ if a change of state happens. As a result, there is a distribution of ages among the population. In the following, we discuss the behavior of the age distributions for the possible states, $x^s_\tau$. They are obtained replacing in Eqs.~\eqref{eq:xxx_tau} the stable steady state values $x^s$, with $x^s=x^s_{\text{\tiny A}}$ if $a<a_\text{c}$, $x^s=x^s_{\text{\tiny S}}$ otherwise. The necessary function $F_\tau(x)$ is given by
\begin{equation}
F_\tau(x)=\gamma(x,a)^\tau,
\end{equation}
where $\gamma(x,a)$ has been defined in Eq.~\eqref{eq:gamma_def}. With this in mind, the normalized age distributions within each population $p^s_\tau=\dfrac{x^s_\tau}{x^s}$ follow a geometric distribution $p^s_\tau=~(1-\Lambda^s)(\Lambda^s)^\tau$ where: \\[10pt]
For $a<a_\text{c}$:
\begin{eqnarray} \label{eq:agingless_lambda_1}
\Lambda^{\pm}&=&\frac{1}{6}\left[3+2a\pm\sqrt{3(1-4a)(3-4a)}\right], \nonumber\\
\Lambda^0&=&\frac{4}{3}a,
\end{eqnarray}
while for $a<a_\text{c}$:
\begin{eqnarray} \label{eq:agingless_lambda_2}
\Lambda^+&=&\Lambda^-=\frac{2-a}{3},\nonumber\\
\Lambda^0&=&\frac{1}{3}.
\end{eqnarray}
From these expressions we can compute the average age of agents in each state 
\begin{equation} \label{eq:taus_agingless}
 \langle \tau^s\rangle=\sum_{\tau=0}^\infty\tau\, p^s_\tau=\frac{\Lambda^s}{1-\Lambda^s},
\end{equation}
which are displayed in Fig.~\ref{fig:theo} (first column, second row). 
It appears from this figure that the $s=+1$ state is not just the most populated for $a<a_\text{c}$ but the average age of agents in that state is also larger than that of the other states. Note that, although for $a<a_\text{c}$ it is $x^-_{\text{\tiny A}}< x^0_{\text{\tiny A}}$, the average times satisfy the reverse inequality $\langle\tau^-\rangle>\langle\tau^0\rangle$, proving that not always the most populated state is the one that has older agents. The same conclusion can be obtained for $a\geq a_\text{c}$, where despite all occupancy fractions being equal, namely $x^\pm_{\text{\tiny S}}=x^0_{\text{\tiny S}}=1/3$, only $\langle\tau^+\rangle$ and $\langle\tau^-\rangle$ take the same values in this noise interval, while $\langle\tau^0\rangle$ is smaller than the common value $\langle\tau^\pm\rangle$. This phenomenon can be attributed to the asymmetry in the kinetic rule described by Eq.~\eqref{eq:kinetic_rule}, which evidences that it is more probable to change the state to $\pm1$ and therefore $\langle\tau^0\rangle$ is smaller. For the limiting case $a=1$, where only random changes can occur, all age distributions for the three possible states, and hence its average value, coincide.

The overall average time $\langle\tau\rangle=\sum_{s=+,0,-}x^s\langle \tau^s\rangle$ is given by 
\begin{equation}
 \langle\tau\rangle= \begin{cases}
 {\displaystyle
\frac{44 a^4-266 a^3+357 a^2-171
 a+27}{a(1-a) (3-4 a) (9-11
 a)}
 }, &a\le a_\text{c}, \\
 {\displaystyle 
 \frac{3-a}{2(1+a)}
 }, &a\ge a_\text{c},
 \end{cases}
\end{equation}
which is also plotted in Fig.~\ref{fig:theo} (first column, second row). 
As it can be seen from this figure, the overall average time obeys, for $a<a_\text{c}$, the inequality $\langle\tau^0\rangle<\langle\tau^-\rangle<\langle\tau\rangle<\langle\tau^+\rangle$, except for a very small range below $a_\text{c}$ where $\langle\tau\rangle<\langle\tau^-\rangle$. For $a>a_\text{c}$, the order is $\langle\tau^0\rangle<\langle\tau\rangle<\langle \tau^\pm\rangle$. 

%%%%%%%%%%%%%%%%%%%%%%%%%%%%%% 

\subsection{Algebraic aging} \label{sec_aging_alg}

In this section we consider that the age-dependent update probability decreases with $\tau$ following an 
algebraic decay law, as 
 previously considered in the literature for the voter model~\cite{Artime2018,Baron2022}.
The precise form is given by
\begin{equation}\label{eq:qalg}
q_\tau=\frac{1}{1+\tau/\tau^*}.
\end{equation} 
The case insensitive to aging is recovered when $\tau^*\to \infty$, in which case 
$q_\tau=1$ for all $\tau$.

The function $\Phi(x)$ for this particular choice of the age profile is given by Eqs.~(\ref{eq:app:sumF},\ref{eq:app:sumqF}) in Appendix~\ref{App_gen_alg}, setting $q_\infty=0,\,q_0=1$.
Although in this case it is not possible to find a closed expression for the solutions of Eqs.~(\ref{dxpdt}-\ref{eq:Fxy}), we have determined them numerically with a very high precision. 
As an example, we plot in Fig.~\ref{fig:theo} (second column, top row) the phase diagram corresponding to $\tau^*=2.0$. We find that the phase diagram is qualitatively similar to that of the aging-less case: The symmetric solution $x^{\pm}_\text{\tiny S}$ always exists and it is stable for $a>a_\text{c}(\tau^*)$. For $a<a_\text{c}(\tau^*)$, the symmetric solution is unstable and a pair of asymmetric stable solutions $x^{+}_\text{\tiny A}$, $x^{-}_\text{\tiny A}$ (corresponding to solutions (II) and (III) of the aging-less case) emerge. The critical value $a_\text{c}(\tau^*)$ is determined by the condition that one of the eigenvalues of the Jacobian matrix, Eq.~\eqref{jacobian}, evaluated at $x^{+}_\text{\tiny S}$ crosses zero (the other eigenvalue turns out to be always negative). In contrast to the aging-less case, for $a>a_\text{c}$, $x^{\pm}_\text{\tiny S}>x^{0}_\text{\tiny S}$, except for $a=1$, where they coincide. Moreover, there is a small region for $a\lesssim a_\text{c}$ for which $x^{-}_\text{\tiny A}>x^{0}_\text{\tiny A}$. 

In Fig.~\ref{fig:simul} (first column) we plot the magnetization (top row), and the fraction $x^0$ of agents in state $0$ (central row), for different values of $\tau^*$, as a function of the noise intensity $a$. Lines correspond to the theoretical results that we just described, and symbols to numerical simulations of the agent-based dynamics, using the stochastic rules of the process in systems of finite population $N$. We can observe a good agreement between theory and simulations, although the latter are naturally affected by finite size effects. We observe a non-monotonic behavior in the magnetization plot: the critical value $a_\text{c}$ tends to $0$ as $\tau^*$ tends to zero, then increases for increasing $\tau^*$ until it reaches $a_\text{c}>1/4$ for intermediate values of $\tau^*$ before tending back to the aging-less value $a_\text{c}=1/4$ for $\tau^*\to\infty$.

As for the aging-less case, one can calculate the average age of the agents in each state $\langle\tau^s\rangle$, setting $q_{\infty}=0,q_0=1$ in Eqs.~(\ref{eq:app:general_taus}, \ref{eq:app:general_T}) of Appendix~\ref{App_gen_alg}. These average aging times $\langle \tau^\pm\rangle$, $\langle \tau^0\rangle$ and the global average $\langle \tau\rangle$, are plotted as a function of $a$, in 
Fig.~\ref{fig:theo} (second column, bottom row).
Compared to the aging-less case (shown in the first column of Fig.~\ref{fig:theo}), we note the same 
inequality relations between the ages of each state. Although $\langle\tau\rangle$ and $\langle\tau^+\rangle$ behave similarly to the aging-less case, the time averages 
$\langle\tau^-\rangle$ and $\langle\tau^0\rangle$ are more sensitive to $\tau^*$. Finally, we note that while $\langle \tau^+\rangle$, and hence $\langle \tau\rangle$, always diverge as $a\to0$, it is found that both $\langle \tau^-\rangle$ and $\langle \tau^0\rangle$ only diverge as $a\to0$ for values of $\tau^*\lesssim 1$, while they remain finite or tend to zero in the same limit otherwise (result not shown).

In Fig.~\ref{fig:simul} (first column, lower row) we show the global average time $\langle \tau\rangle$ vs. $a$ for different values of $\tau^*$. The figure shows again the good agreement of the mean-field theory with simulations of the agent-based model. Moreover note that for any value of $a$ the average age increases the stronger the aging effect is (the smaller $\tau^*$), both in the ordered and disordered regimes.

\begin{figure}[h!]
\centering
\includegraphics[width = 0.95\columnwidth]{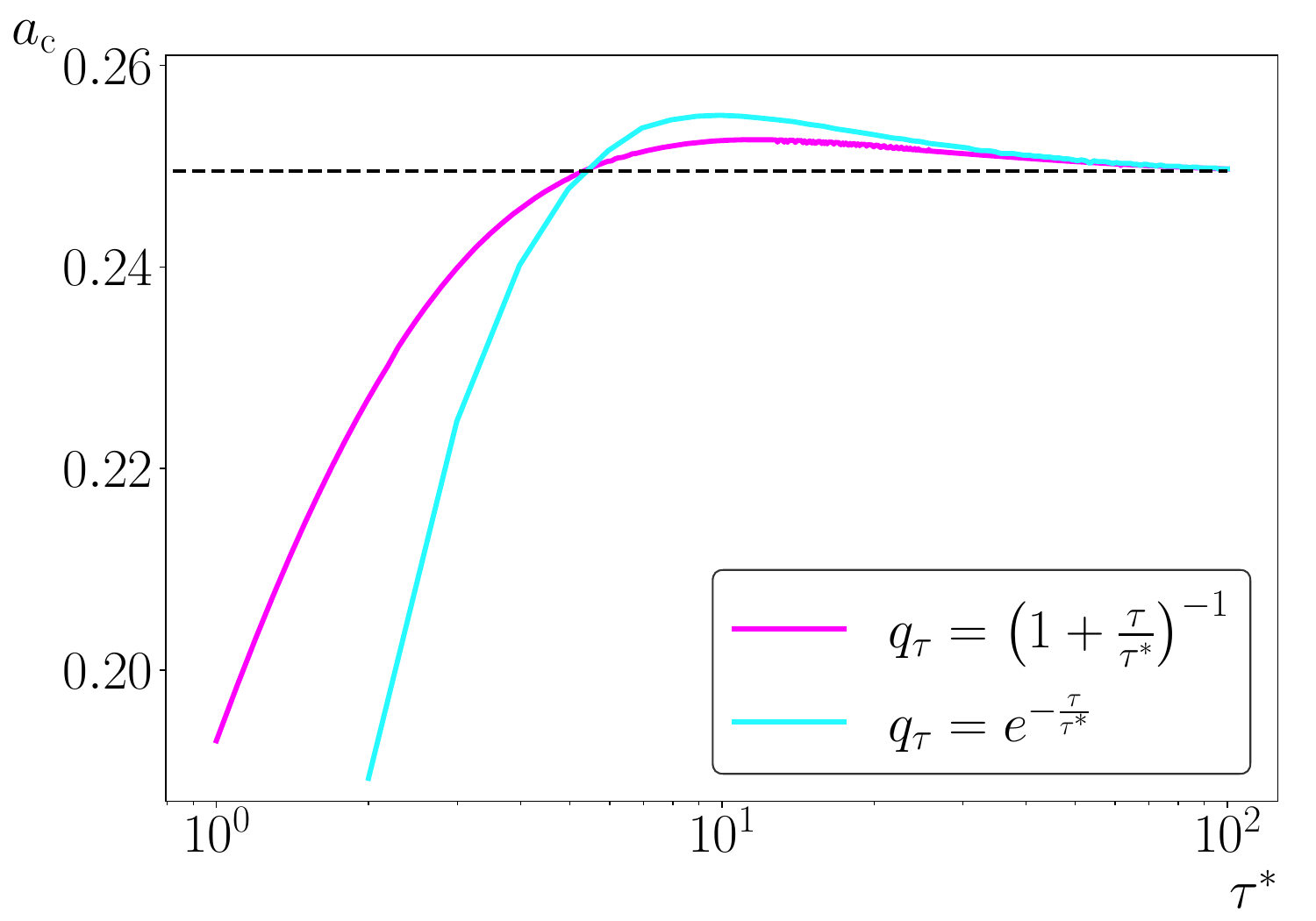}
 \caption{Critical curve $a_\text{c}$ vs. $\tau^*$. The solid lines are the theoretical results. 
 The dashed horizontal line represents the aging-less case, $a_\text{c}=1/4$, drawn for comparison. We plot results both for the algebraic, Eq.~\eqref{eq:qalg}, and the exponential, Eq.~\eqref{eq:q_exp}, dependence of the aging update probability. Note that there is a regime of parameter values $\tau^*>\tau^*_c$ where $a_\text{c}>1/4$, indicating that the phase diagram in the presence of aging shows a larger area of validity of the non-symmetric solution compared to the non-aging case. The contrary occurs for $\tau^*<\tau^*_c$. In the algebraic case, it is $\tau^*_c=5.6171828\dots$, while the exponential case leads to $\tau^*_c=5.39971436\dots$}
 \label{fig:ac-vs-tau}
\end{figure}

Finally, we have computed numerically the curve $a_\text{c}$ vs. $\tau^*$ and plotted it in Fig.~\ref{fig:ac-vs-tau}. This plot evidences the non-monotonic behavior observed in Fig.~\ref{fig:simul}. For $\tau^*<\tau^*_c$, aging hampers consensus with respect to the aging-less while, for $\tau^*>\tau^*_c$, the ordered phase becomes wider. The critical value $\tau^*_c=5.6171828\dots$ is determined numerically. In the limit $\tau^*\to \infty$, we recover the value $a_\text{c}=1/4$ as expected. We can infer that strong aging, low $\tau^*$, obstructs the formation of a consensus. However, when appropriately regulated by the value of $\tau^*>\tau^*_c$, such aging can contribute to favor the consensus formation. 

%%%%%%%%%%%%%%%%%%%%%%%%%%%%%%
\subsection{Exponential aging}

Now, we consider an exponential aging profile 
\begin{equation}\label{eq:q_exp}
q_\tau=\exp(-\tau/\tau^*),
\end{equation} 
 which has been previously considered for the noisy voter model~\cite{Baron2022}. 
This law provides a much faster decrease of the probability of using the influence rule as a function of the age the agent has spent in its current state. As in the case of an algebraic dependence, the lack of influence of aging (i.e., $q_\tau=1$ for all $\tau\ge 0$) is recovered when $\tau^*\to \infty$. Unfortunately, we have not been able to find an analytical expression for $\Phi(x)$. Nevertheless, the values of this function can be determined very precisely by an efficient numerical routine. The details of the calculation of $\Phi(x)$ are relegated to the Appendix~\ref{app_exponential}.

In general terms, the exponential case presents qualitative similarities with the algebraic case, as shown in both phase diagram and ages distribution of Fig.~\ref{fig:theo} (third column). The main difference is that $\langle \tau^-\rangle$ and $\langle \tau^0\rangle$ vanish for any $\tau^*$ in the limit $a\to 0$. Moreover, Fig.~\ref{fig:simul} (third column) shows excellent agreement between theory and simulations. 
 
 The non-monotonic behavior of the magnetization appears again. 
 We have also determined numerically the curve $a_\text{c}$ vs. $\tau^*$, displayed in Fig.~\ref{fig:ac-vs-tau}, which qualitatively resembles that of the algebraic case. Since the exponential kernel decays with $\tau^*$ faster than the algebraic kernel, the effect of aging is stronger and the interval of $a$ associated to the ordered phase is narrower when $\tau^*<\tau^*_c$. However, for $\tau^*>\tau^*_c$ the system presents a wider ordered region. The critical value $\tau^*_c=5.39971436\dots$ has been determined numerically. Therefore, there are optimal values of $\tau^*$ for which consensus is favored even more than in the algebraic case. For $\tau^*$ large enough, both exponential and algebraic cases coincide and in the limit $\tau^*\to\infty$, $a_\text{c}=1/4$.

%%%%%%%%%%%%%%%%%%%%%%%%%%%%%%
\subsection{Anti-aging}\label{sec_anti_alg}
In addition to the aging scenario considered so far, we address in this section the situation in which it is more likely to interact with the neighbors the longer the persistence in the current state. This propensity to change with the age can be modeled by a factor $q_\tau$ that is an increasing function of $\tau$. Namely, we adopt the expression
\begin{equation}\label{eq:q_anti}
q_\tau=\frac{q_0+ \tau/\tau^*}{1+ \tau/\tau^*},
\end{equation}
with $\tau^*>0$ and $0<q_0=q(\tau=0)\le 1$. This functional form has been previously considered in the context of the noisy voter model~\cite{Peralta2020a,Baron2022}. The aging-less case is recovered for $q_0=1$.

The function $\Phi(x)$ associated to Eq.~(\ref{eq:q_anti}) is obtained by setting $q_\infty=~1$ in Eqs.~(\ref{eq:app:sumF}-\ref{eq:app:sumqF}) of Appendix~\ref{App_gen_alg}. The theoretical phase diagram for the anti-aging case is illustrated in Fig.~\ref{fig:theo} (fourth column), for $\tau^*= 0.1 $ and $q_0=0.1$. Note that, in contrast to the studied cases with aging, here the neutral fraction $x^0$ dominates in the disordered phase, exceeding the value 1/3. In Fig.~\ref{fig:simul} (third column), we plot the order parameter $m$ and the fraction $x^0$ vs. $a$, for $\tau^*=0.1$ and different values of $q_0$, compared to simulations.

The analytical expression for the average age of the agents in each state $\langle\tau^s\rangle$ is obtained by setting $q_{\infty}=0,q_0=1$ in Eqs.~(\ref{eq:app:general_taus}, \ref{eq:app:general_T}) of Appendix~\ref{App_gen_alg}. These average aging times $\langle \tau^\pm\rangle$, $\langle \tau^0\rangle$ and the global average $\langle \tau\rangle$, are plotted as a function of $a$ in Fig.~\ref{fig:theo} (fourth column, bottom row). In addition, we compare in the fourth column of Fig.~\ref{fig:simul} these theoretical expressions of the average aging with the results coming from numerical simulations, showing an excellent agreement between both. For fixed $\tau^*$, a decrease of $q_0$ (i.e., the probability $q(\tau=0)$) reduces $a_\text{c}$ monotonically, while for any value of $a$ it enhances the population of neutral individuals given by $x^0$, and reduces the average time $\langle \tau \rangle$ in the ordered phase.

\begin{figure}[h]
\includegraphics[width = 0.95\columnwidth]{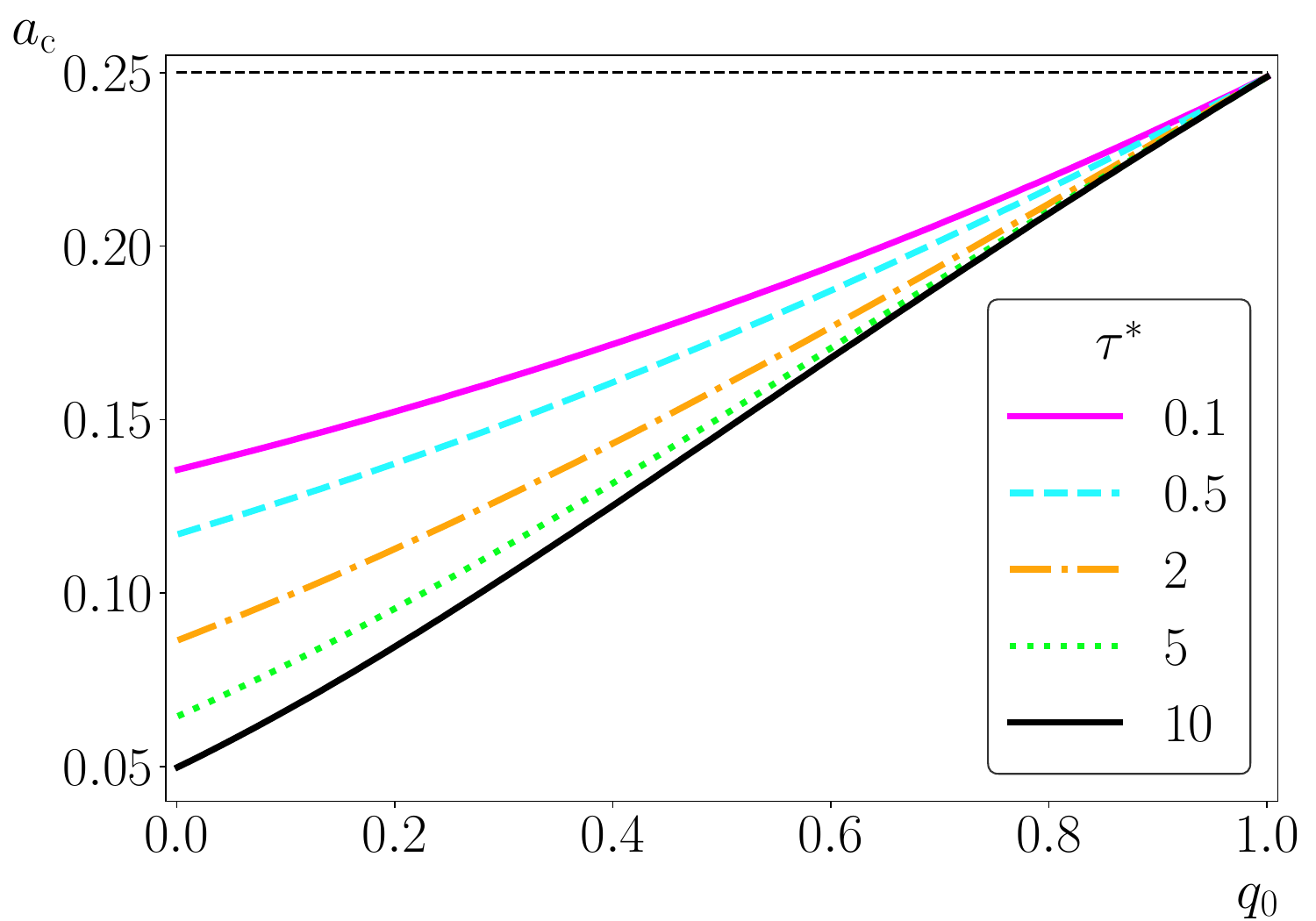}
 \caption{Critical value of the noise intensity $a_\text{c}$ vs. $q_0$ for fixed values of $\tau^*$, for the anti-aging scenario given by 
Eq.~\eqref{eq:q_anti}. The horizontal dashed line indicates the critical value $a_\text{c}=1/4$ in the aging-less case.}
 \label{fig:critical_anti}
\end{figure}

As a summary, we can conclude that the use of the anti-aging profile of Eq.~\eqref{eq:q_anti}, for any value of the parameters $q_0$ and $\tau^*$, contributes to reduce the critical value $a_\text{c}$, compared to the aging-insensitive case. This is evidenced in the plot of $a_\text{c}$ versus $q_0$ for several values of $\tau^*$ displayed in Fig.~\ref{fig:critical_anti}. 

%%%%%%%%%%%%%%%%
%%%%%%%%%%%%%%%%%%%%%%%%%%%%
\section{Final remarks}
\label{sec:final}

We have considered a 3-state kinetic model for opinion formation and introduced the effect of aging in the dependence $q_\tau$ of the probability to change state when an interaction between agents occurs. We have derived a mean-field description and introduced an adiabatic approximation in order to derive evolution equations for the density of agents in each opinion state. We have analyzed the phase diagram using the steady state solutions and their stability. The theory works for arbitrary forms of the aging factor $q_\tau$ and we have validated it by comparing its results with numerical simulations of the stochastic agent-based model in a complete graph. 

We have considered particularly the algebraic, $q_\tau = 1/(1+ \tau/\tau^*)$, and exponential, $q_\tau = \exp(-\tau/\tau^*)$, functional forms for $q_\tau$. Our results indicate that qualitatively similar phenomena emerge for both cases. We have also considered numerical simulations using the generalized algebraic form $q_\tau = 1/(1+ \tau/\tau^*)^\alpha$, with $\alpha>0$, and did not observe any new qualitative features with respect to the case $\alpha=1$ (results not shown). For very large $\tau^*$ (weak aging), there is a good agreement with the aging-less situation, as expected. For all values of $\tau^*$, the magnetization $m$ as a function of the noise-intensity $a$ displays a continuous transition from order to disorder at a critical value $a_\text{c}$. For weak aging (i.e., large $\tau^*$), increasing aging behavior (that is, decreasing $\tau*$) favors order, shifting the critical value $a_\text{c}$ to larger values, a qualitatively similar trend to that observed in the voter model (in which case $a_\text{c}=0$ in absence of aging, differently to the 3-state model here considered). 
However, for sufficiently strong aging (i.e., small $\tau^*$), the opposite trend is observed with a reduction of the critical value by increasing the aging behavior.
This indicates that the critical point has a non-monotonous dependency on the characteristic time $\tau^*$, such that there is an optimal value of $\tau^*$ that favors order, as observed in Fig.~\ref{fig:ac-vs-tau}. We also considered an anti-aging scenario given by $q_\tau=(q_0+\tau/\tau^*)/(1+\tau/\tau^*)$, in which case we observe a reduction of the region of consensus for any value of the parameters $q_0$ and $\tau^*$.
 
As perspectives of continuation, it would be interesting to consider other ways in which the aging can affect the updating rules. For example, aging could have an impact on the updating probabilities of the random changes as well as on those due to interactions. It would also be interesting to test the robustness of the observed effects by considering other variants of this kinetic exchange opinion model, as well as to go beyond the simplistic all-to-all interactions and to study the model on random networks. 

\acknowledgments{
Partial financial support has been received from the Agencia Estatal de Investigaci\'on (AEI, MCI, Spain) MCIN/AEI/10.13039/501100011033 and Fondo Europeo de Desarrollo Regional (FEDER, UE) under Project APASOS (PID2021-122256NB-C21) and the María de Maeztu Program for units of Excellence in R\&D, grant CEX2021-001164-M. 
C.A. also acknowledges partial support received from
 Conselho Nacional de Desenvolvimento Científico e Tecnológico (CNPq)-Brazil (311435/2020-3) and Fundação de Amparo à Pesquisa do
Estado de Rio de Janeiro (FAPERJ)-Brazil (CNE E-26/201.109/2021). 
}

\bibliography{Aging}

%%%%%%%%%%%%%%%%%%%%%%%%%%%%%% App
%%%%%%%%%%%%%%%%%%%%%%%%%%%%%% App
%\cleardoublepage
%\newpage
\onecolumngrid
\appendix
\section{Global rates}\label{app_xtau}
Let us suppose that, at some time $t$, $n_{\tau}^-$
agents are in the state $-1$ and have age $\tau$. If we randomly select one of these agents, it can face the following situations:
\begin{enumerate}
 \item Remain at the same state $s=-1$, i.e., $n^{-}_\tau \to n^{-}_{\tau}-1 ,\, n^{-}_{\tau+1} \to n^{-}_{\tau+1}+1$, which can occur by
 \begin{itemize}
 
 \item random choice with probability $a/3$;
 
 \item interacting with any of the $n^-$ agents that are in state $-1$, with probability $\frac{n^-}{N}(1-a)$, either the kinetic rule is applied, with probability $q(\tau)$, or not, with probability $1-q(\tau)$; 
 
 \item interacting with any of the $n^0$ agents that are in state $0$, with probability $\frac{n^0}{N}(1-a)$, either the kinetic rule is applied, with probability $q(\tau)$, or not, with probability $1-q(\tau)$; 
 
 \item interacting with any of the $n^+$ agents that are in state $+1$ without following the kinetic rule, Eq.~\eqref{eq:kinetic_rule}, with probability $\frac{n^+}{N}(1-a)\left(1-q(\tau)\right)$. 
 
 \end{itemize}
 
 The global rate taking into account these possible transitions reads
 \begin{equation}
 \Omega_1(\tau)=n_{\tau}^{-}\left(\frac{a}{3}+(1-a)\left[1-\left(1-x^0-x^-\right)q_{\tau}\right]\right),
 \end{equation}
 where for brevity in the notation we write $q_{\tau}=q(\tau)$ and $n^{s}_\tau$ is the number of agents with opinion $s$ and age $\tau$, $x^s_\tau= n^s_\tau/N$, 
$x^s = \sum^\infty_{\tau=0} x^s_\tau$.

 \item Change to state $s=0$, i.e., $n^{-}_\tau \to n^{-}_{\tau}-1 ,\, n^{0}_{0} \to n^{0}_{0}+1$, which can occur by
 \begin{itemize}
 \item random choice with probability $a/3$, 
 
 \item interacting with any of the $n^+$ agents that are in state $+1$ without applying the kinetic rule, with probability $\frac{n^+}{N}(1-a)\left(1-q_{\tau}\right)$.
 \end{itemize}
 
 The global rate taking into account these possible transitions reads
 \begin{equation}
 \Omega_2(\tau)=n_{\tau}^{-}\left(\frac{a}{3}+(1-a)x^+q_{\tau}\right),
 \end{equation}
 \item Finally, change to state $s=+1$, i.e., $n^{-}_\tau \to n^{-}_{\tau}-1 ,\, n^{+}_{0} \to n^{+}_{0}+1$, situation only possible by random change with probability $a/3$. In this case the global rate is simply
 \begin{equation}
 \Omega_3(\tau)=n_{\tau}^{-}\;\frac{a}{3}.
 \end{equation}
\end{enumerate}

The other global rates shown in Table~\ref{app:tab:global_rates} are calculated in a similar way. From these global rates, it is easy to obtain that
\begin{eqnarray}
 \Omega_1 (\tau) + \Omega_2 (\tau) + \Omega_3 (\tau) = n^{-}_\tau, \\
\Omega_4 (\tau) + \Omega_5 (\tau) + \Omega_6 (\tau) = n^{0}_\tau, \\
\Omega_7 (\tau) + \Omega_8 (\tau) + \Omega_9 (\tau) = n^{+}_\tau,
\end{eqnarray}
 
which leads to the evolution equations:
\begin{itemize}
 \item For $\tau\ge 0$:
\begin{eqnarray}
\frac{d n^{-}_\tau}{dt} &=& \Omega_1 (\tau - 1) - \Omega_1 (\tau) - \Omega_2 (\tau) - \Omega_3 (\tau) ,\\
\frac{d n^{0}_\tau}{dt} &=& - \Omega_4 (\tau) - \Omega_5 (\tau) + \Omega_5 (\tau - 1) - \Omega_6 (\tau) ,\\
\frac{dn^{+}_\tau}{dt} &=& - \Omega_7 (\tau) - \Omega_8 (\tau) + \Omega_9 (\tau- 1) - \Omega_9 (\tau).
\end{eqnarray}
\item For $\tau=0$,
 \begin{eqnarray}
 \frac{d n^{-}_0}{dt} &=& \sum_{\tau=0}^{\infty} \left[ \Omega_4 (\tau) + \Omega_7 (\tau) \right] - \
\Omega_1 (0) - \Omega_2 (0) - \Omega_3 (0) = \sum_{\tau=0}^{\infty} \left[
\Omega_4 (\tau) + \Omega_7 (\tau)\right] - n^{-}_0, \\
\frac{d n^{0}_0}{dt} &=& \sum_{\tau=0}^{\infty} \left[\Omega_2 (\tau) + \Omega_8 (\tau) \right] - \
\Omega_4 (0) - \Omega_5 (0) - \Omega_6 (0) =\sum_{\tau=0}^{\infty} \left[ \
\Omega_2 (\tau) + \Omega_8 (\tau) \right] - n^{0}_0, \\
\frac{d n^{+}_0}{dt} &=& \sum_{\tau=0}^{\infty} \left[ \Omega_3 (\tau) + \Omega_6 (\tau) \right]- \
\Omega_7 (0) - \Omega_8 (0) - \Omega_9 (0) =\sum_{\tau=0}^{\infty} \left[ \
 \Omega_3 (\tau) + \Omega_6 (\tau) \right] - n^{+}_0.
\end{eqnarray}
\end{itemize}
After replacing the rates of Table~\ref{app:tab:global_rates}, we obtain Eqs.\~(\ref{eq:ratesi1},\ref{eq:ratesi0}) in the main text for the densities $x_\tau^s=\dfrac{n_\tau^s}{N}$.

\begin{table}[h]
\renewcommand{\arraystretch}{1.5}
\centering
\begin{tabular}{|c|cc|l|}
\hline
$s(t) \to s(t+1)$ & \multicolumn{2}{c|}{Transition} & Global rate \\ \hline
$-1 \to -1$ & \multicolumn{1}{l|}{$n^{-}_{\tau+1} \to n^{-}_{\tau+1}+1$} & \multirow{3}{*}{$n^{-}_\tau \to n^{-}_{\tau}-1$} & $\Omega_1 (\tau)=n^-_\tau\gamma(x^+q_\tau,a)$
\\ \cline{1-2} \cline{4-4} 
$ -1 \to 0$ & \multicolumn{1}{l|}{$n^0_{0} \to n^0_{0} +1 $} & & $\Omega_2 (\tau)=n^-_\tau\gamma(1-x^+q_\tau,a)$
\\ \cline{1-2} \cline{4-4} 
$-1 \to 1$ & \multicolumn{1}{l|}{$n^{+}_{0} \to n^{+}_{0} +1 $} & & $\Omega_3 (\tau)=n^-_\tau\gamma(1,a)$
\\ \hline
$0 \to -1$ & \multicolumn{1}{l|}{$n^{-}_{0} \to n^{-}_{0}+1$} & \multirow{3}{*}{$n^0_\tau \to n^0_\tau-1$} & $\Omega_4 (\tau) = n^0_\tau\gamma(1-x^-q_\tau,a)$
\\ \cline{1-2} \cline{4-4} 
$0 \to 0$ & \multicolumn{1}{l|}{$n^0_{\tau+1} \to n^0_{\tau+1} +1$} & & $\Omega_5 (\tau) =n^0_\tau\gamma((1-x^0)q_\tau,a)$
\\ \cline{1-2} \cline{4-4} 
$0 \to +1$ & \multicolumn{1}{l|}{$n^{+}_{0} \to n^{+}_{0}+1$} & & $\Omega_6 (\tau) = n^0_\tau\gamma(1-x^+q_\tau,a)$
\\ \hline
$+1 \to -1$ & \multicolumn{1}{l|}{$n^{-}_{0} \to n^{-}_{0}+1$} & \multirow{3}{*}{$n^{+}_\tau \to n^{+}_\tau-1$} & $ \Omega_7 (\tau) =n^+_\tau\gamma(1,a)$
\\ \cline{1-2} \cline{4-4} 
$+1 \to 0$ & \multicolumn{1}{l|}{$n^0_{0} \to n^0_{0} +1$} & & $\Omega_8 (\tau) = n^+_\tau\gamma(1-x^-q_\tau,a)$
\\ \cline{1-2} \cline{4-4} 
$+1 \to +1$ & \multicolumn{1}{l|}{$n^{+}_{\tau+1} \to n^{+}_{\tau+1}+1$} & & $\Omega_9 (\tau)=n^+_\tau\gamma(x^-q_\tau,a)$
\\ \hline
\end{tabular}
\caption{ Global rates of the process, recalling that $\gamma(z,a)\equiv\frac{a}{3}+(1-a)(1-z)$, as defined in Eq.~(\ref{eq:gamma_def}).} 
%$\alpha_\tau(x) = a/3 + (1 - a) [1 - (1-x )q_\tau]$ and 
%$\beta_\tau(x) = a/3 + (1 - a)\, x \,q_\tau$.
\label{app:tab:global_rates}
\end{table}
%%%%%%%%%%%%%%%%%%%%%%%%%%%%%% App
\section{Calculation of the function $\Phi(x)$ for a general rational function of the age}\label{App_gen_alg}

We consider the general rational functional form proposed in~\cite{Peralta2020a} 
\begin{equation}\label{eq:app:q_general}
 q_{\tau}=\frac{q_{\infty}\tau+q_0\tau^*}{\tau+\tau^*},
\end{equation}
where $q_{\infty},\,q_0\in[0,1]$ and $\tau^*>0$. If $q_0>q_{\infty}$, $q_{\tau}$ is a decreasing function with the age, i.e., agents are less likely to copy their neighbors. This is a typical aging situation. The opposite occurs if $q_0<q_{\infty}$, where $q_{\tau}$ increases with the age. This is an anti-aging situation. 

Using the definition Eq.~(\ref{eq:Ftau}) and Eq.~(\ref{eq:app:q_general}), the function $F_{\tau}(x)$ becomes
\begin{equation}
 F_{\tau}(x)=\gamma(q_\infty\,x,a)^{\tau }\frac{ \left(\tau^*\xi(x,q_0,q_\infty,a)\right)_{\tau }}{(\tau^*)_{\tau }},
\end{equation}
with
\begin{equation}
 \xi(x,q_0,q_\infty,a)\equiv\frac{\gamma(q_0x,a)}{\gamma(q_\infty x,a)},
\end{equation}
where
$(z)_\tau\equiv \Gamma(z+\tau)/\Gamma(z)$ is the Pochhammer symbol, and $\gamma(z,a)$ is given in Eq.~\eqref{eq:gamma_def}.

In order to compute the function $\Phi(x)=\displaystyle\frac{\sum_{\tau=0}^\infty q_\tau F_\tau(x)}{\sum_{\tau=0}^\infty F_\tau(x)}$, we need the following sums: 
\begin{align} \label{eq:app:sumF}
 \sum_{\tau=0}^{\infty}F_{\tau}(x)=&\,_2F_1\left(1,\tau^*\xi(x,q_0,q_\infty,a);\tau^*;\gamma(q_\infty\,x,a) \right),\\
 \sum_{\tau=0}^{\infty}q_{\tau}F_{\tau}(x)=&q_0 \,_2F_1\left(1,\tau^*\xi(x,q_0,q_\infty,a);1+\tau^*;\gamma(q_\infty\,x,a) \right)+ \nonumber\\
 &\frac{q_{\infty} }{1+\tau^*}\gamma(q_0\,x,a)\, _2F_1\left(2,1+\tau^*\xi(x,q_0,q_\infty,a);2+\tau^*;\gamma(q_\infty\,x,a) \right). \label{eq:app:sumqF}
\end{align}

The convergence of the hypergeometric functions $_2F_1$ appearing in these expressions is guaranteed as $|\gamma(z,a)|<1$ for all values of the parameters and variables that appear in them. There are very efficient routines to compute numerically the hypergeometric functions and, hence, the function $\Phi(x)$ needed for the determination of the phase diagrams. In this work we have used the implementation in Mathematica and in the SciPy Python package.

One can also calculate the function 
\begin{equation}
 \sum_{\tau=0}^{\infty}\tau F_{\tau}(x)=\gamma(q_0x,a)\,_2F_1\left(2,1+\tau^*\xi(x,q_0,q_\infty,a);1+\tau^*;\gamma(q_\infty\,x,a) \right),
\end{equation}
which is needed for calculating the average age of agents in each state $\langle\tau^s\rangle$, which is given by \begin{equation}
\langle\tau^s\rangle=\frac{1}{{x^s}}\sum_{\tau}\tau\, x^s_{\tau},
\end{equation}
where $x^s_{\tau}$ is defined in Eqs.~\eqref{eq:xxx_tau}. In this way, one obtains
\begin{equation}\label{eq:app:general_taus}
 \begin{split} 
\langle\tau^{\pm}\rangle=\mathcal{T}\left(x^{\mp},q_0,q_{\infty},a\right),\thinspace 
 \langle\tau^{0}\rangle=\mathcal{T}\left(1-x^0,q_0,q_{\infty},a\right), 
 \end{split}
\end{equation}
where $x^s,\,s=-1,0,+1$ are the stable solutions of Eqs.~(\ref{dxpdt}-\ref{eq:Fxy}) and the function $\mathcal{T}$ is given by
\begin{equation} \label{eq:app:general_T}
\mathcal{T}\left(x,q_0,q_{\infty},a\right)=\gamma(q_0x,a)\,\frac{_2F_1\left(2,1+\tau^*\xi(x,q_0,q_\infty,a);1+\tau^*;\gamma(q_\infty\,x,a)\right)}{_2F_1\left(1,\tau^*\xi(x,q_0,q_\infty,a);\tau^*;\gamma(q_\infty\,x,a) \right)}.
\end{equation}
From these general expressions we can recover the different cases analyzed in the main text. The algebraic aging of section \ref{sec_aging_alg} corresponds to setting $q_{\infty}=0$, $q_0=1$. The anti-aging case considered in section \ref{sec_anti_alg} corresponds to the case $q_{\infty}=1$ and $q_0$ variable. Finally, even the aging-less case of section \ref{sec:aging-less} is recovered by setting $q_{\infty}=q_0=1$ that leads to $q_\tau=1$. Since $\xi(x,1,1,a)=1$ one can write Eq.~\eqref{eq:app:general_T} as
\begin{equation} \label{eq:app:taus_agingless}
\mathcal{T}\left(x,1, 1,a\right)=\gamma(x,a)\,\frac{\sum_{n=0}^{\infty}(2)_n\frac{\gamma(x,a)^n}{n!}}{\sum_{n=0}^{\infty}(1)_n\frac{\gamma(x,a)^n}{n!}}=\frac{\gamma(x,a)}{1-\gamma(x,a)},
\end{equation}
which is precisely Eq.~\eqref{eq:taus_agingless}. Eqs.~(\ref{eq:agingless_lambda_1}, \ref{eq:agingless_lambda_2}) can be obtained substituting Eqs.~(\ref{eq:S},\ref{eq:A}) into Eqs.~(\ref{eq:app:general_taus}, \ref{eq:app:taus_agingless}).
%%%%%%%%%%%%%%%%%%%%%%%%%%%%

\section{Calculation of the function $\Phi(x)$ for the exponential aging}\label{app_exponential}

Using $q_\tau=e^{-\tau/\tau^*}$, we obtain from the definition given by Eq.~\eqref{eq:Ftau} that the function $F_\tau(x)$ is explicitly given by
\begin{equation}
F_\tau(x)=\left(1-\frac{2a}{3}\right)\left(3x\frac{1-a}{3-2a};e^{-1/\tau^*}\right)_{\tau},
\end{equation}
where $(b;s)_0=1$ and $(b;s)_n=(1-b)(1-bs)(1-bs^2)\cdots(1-bs^{n-1})$ for $n\ge 1$, is the $q$-Pochhammer symbol. 
The desired function $\Phi(x)$ is given by the ratio
\begin{equation}\label{eq:Phi_sum}
\Phi(x)=\frac{\sum_{\tau=0}^\infty q_\tau F_\tau(x)}{\sum_{\tau=0}^\infty F_\tau(x)}=\frac{\phi(\alpha s,b,s)}{\phi(\alpha,b,s)},
\end{equation}
where 
\begin{equation}\label{eq:phi}
\phi(\alpha,b,s)=\sum_{\tau=0}^\infty (b;s)_\tau\,\alpha^\tau 
\end{equation}
and
\begin{equation}
\begin{split}
b&=3x\frac{1-a}{3-2a}\in(0,1),\\
s&=e^{-1/\tau^*}\in(0,1),\\
\alpha&=1-\frac{2a}{3}\in(1/3,1).
\end{split}
\end{equation}
It does not seem to be possible to express $\phi(\alpha,b,s)$ in terms of other known functions. A possibility to compute numerically the function $\phi(\alpha,b,s)$ is to cut-off the infinite sums in its definition, Eq.~\eqref{eq:phi}, to an upper index $L$ and take $L$ large enough. However, the series seems to be slowly convergent, specially for large values of $\alpha$, and the calculation of each term is costly in terms of computer time. We have found a much more efficient numerical scheme, valid for $s< 1$, to evaluate the function $\phi(\alpha,b,s)$, and hence $\Phi(x)$. This is based on the iteration of the following functional relation than can be easily derived from the definition of $\phi(\alpha,b,s)$ and the properties of the $q$-Pochhammer symbol,
\begin{equation}\label{recursion}
\begin{split}
\phi(\alpha,b,s)&=1+\alpha(1-b)\,\phi(\alpha,bs,s),\\
\phi(\alpha,bs,s)&=1+\alpha(1-bs)\,\phi(\alpha,bs^2,s),\\
\phi(\alpha,bs^2,s)&=1+\alpha(1-bs^2)\,\phi(\alpha,bs^3,s),\\ 
\cdots\cdots\cdots\cdots&\cdots\cdots\cdots\cdots\cdots\cdots\cdots\cdots\cdots\cdots\\
\phi(\alpha,bs^{n-1},s)&=1+\alpha(1-bs^{n-1})\,\phi(\alpha,bs^n,s). 
\end{split}
\end{equation}
If we stop the previous iteration scheme at a sufficiently large value of $n$, we can replace $b s^n$ by zero, using that $s<1$, and then make use of the limiting value
\begin{equation}\label{limiting}
\phi(\alpha,0,s)=\frac{1}{1-\alpha},
\end{equation}
that follows readily from the definition~\eqref{eq:phi} and $(0;s)_n=1$. In practice we have used the criterion to stop the iteration when $bs^n<\epsilon$ with $\epsilon=10^{-8}$, and we have checked that smaller values of $\epsilon$ did not make any difference in the calculation. This provides a very efficient algorithm for the calculation of the function $\phi(\alpha,b,s)$ that requires of the order of 
$n\sim\log(\epsilon/b)/ \log{s}=\tau^*\log(b/\epsilon)$
simple operations (additions and multiplications) and works well for all values of $x,a,\tau^*$.
\end{document}